# Automated-Truck Penetration Shapes Human Drivers' Car-Following Margins Near Motorway Ramps: A Driving-Simulator Study

Minhao Chen[a], Zehao Jiang[b], Julius Secadiningrat[a], Elnaz Irannezhad[a,*]

[a] Research Centre for Integrated Transport Innovation (rCITI), School of Civil and Environmental Engineering, University of New South Wales, Sydney, NSW 2052, Australia

[b] National Centre of Technology Innovation for Digital Construction, School of Civil and Hydraulic Engineering, Huazhong University of Science and Technology, Wuhan 430074, China

*Corresponding author: Elnaz Irannezhad, e.iranenzhad@unsw.edu.au

**Abstract**

Automated-vehicle dedicated lanes (AVDLs) are proposed as transitional infrastructure for mixed traffic, but their effect near ramps, and that of automated trucks, on the human drivers sharing the road has rarely been observed. In a mixed-design driving-simulator experiment, 63 drivers were assigned to one of four layouts (an AVDL extending beyond, coinciding with, or ending inside a 1.5-km ramp weaving section, or no AVDL) and drove at four automated-truck penetration levels (20–80%). Linear mixed-effects models with participant random effects and multiplicity control were used. Higher penetration reduced the share of car-following time at gaps below 1.5 s from 28.5% to 18.1% (Holm-adjusted p = .006). This was explained by leader composition, where drivers spent 8% of following time behind a truck at 20% penetration and 48% at 80%, kept longer gaps behind trucks, and did not change their gaps within leader type. AVDL layout had no detectable effect on safety margins, rear-end conflict rates, travel time or speed (differences bounded within ±5 s per section) and no significant effect on workload. Between-driver differences accounted for 52–77% of the variance in the main outcomes. Mixed-traffic models should therefore represent human car-following as leader-type-dependent and heterogeneous.

**Keywords:** automated vehicle dedicated lane; automated truck; mixed traffic; car-following; driving simulator; driver behaviour

## 1 Introduction

The gradual introduction of automated vehicles (AVs) into motorway networks will create a long transition period in which automated and human-driven vehicles share the same roads. Automated-vehicle dedicated lanes (AVDLs), meaning lanes reserved for eligible automated vehicles, have been proposed as transitional infrastructure that can separate heterogeneous vehicle types, support cooperative operation and improve throughput (Rad et al., 2020; Ghiasi et al., 2017). Corridor-scale dedicated-lane projects are under discussion in several countries (Coyner et al., 2025), and a growing body of optimisation and simulation work addresses where such lanes should be located and how long they should be (Liu and Song, 2019; Kumar et al., 2020; Madadi et al., 2021; Ngoduy et al., 2024; Liu et al., 2025).

The layout problem is most acute in ramp influence areas (RIAs). When an AVDL occupies a motorway lane through a weaving section, automated vehicles must enter and leave the dedicated lane in the same space in which merging and diverging vehicles change lanes. The position of the AVDL entry and exit relative to the ramp gores therefore determines whether these lane-changing processes overlap (Liu et al., 2025). Simulation studies indicate that this choice affects capacity and surrogate safety, but they represent human drivers through fixed model parameters. Whether human drivers actually respond differently to alternative AVDL layouts near ramps has not, to our knowledge, been measured.

A second open question concerns automated platooned trucks. Heavy vehicles already alter surrounding traffic through their length, acceleration and visual occlusion (Al-Kaisy et al., 2002; Peeta et al., 2005), and automated trucks operating as a platoon may amplify these effects. Studies of human drivers near AVs and platoons report behavioural adaptation, such as shorter headways after exposure to short-gap platoons (Gouy et al., 2014) and changes in car-following near dedicated lanes (Schoenmakers et al., 2021; Razmi Rad et al., 2024). When the share of automated trucks increases, two distinct mechanisms could change a human driver's safety margins. Drivers could adapt their car-following policy, or the mix of vehicles they follow could change while their policy stays the same. The two mechanisms have different implications for traffic modelling. The first requires penetration-dependent driver parameters; the second requires leader-type-dependent parameters, which most mixed-traffic models do not include (Talebpour and Mahmassani, 2016; Chen et al., 2017).

A third consideration is heterogeneity among human drivers. Car-following behaviour varies substantially across drivers (Ossen et al., 2006; Ossen and Hoogendoorn, 2011), and a mixed design in which each driver experiences several conditions makes it possible to separate between-driver variation from condition effects. That separation matters both for inference and for the calibration of human driving models used to evaluate AVDL designs.

This study reports a controlled driving-simulator experiment in which 63 licensed drivers were assigned to one of four AVDL layouts and drove through three motorway weaving sections at four automated-truck penetration levels. This paper addresses three research questions:

RQ1: How do AVDL layout and automated-truck penetration affect human drivers' car-following safety margins and rear-end conflicts, and does any penetration effect reflect adaptation of following behaviour or a change in the vehicles being followed?

RQ2: Are AVDL layout and penetration associated with drivers' subjective workload, simulator sickness, heart rate and the distribution of their gaze across vehicles and the road scene?

RQ4: Can human car-following among automated trucks be represented by a car-following model whose parameters depend on leader type, and how large is the leader-type effect compared with differences between drivers?

The study makes three main contributions. First, it provides human-in-the-loop evidence on AVDL layouts near ramps. The evidence covers safety margins, rear-end conflicts, travel time over identical road sections, workload and gaze, and it is analysed with models that respect the mixed design and control for multiple testing. Second, it decomposes the effect of automated-truck penetration on human-driven vehicle safety margins into leader composition and within-leader behaviour, and shows that drivers' gaze shifts between vehicle types in the same way. Third, it calibrates a leader-type-dependent car-following model for human drivers among automated trucks on simulator trajectories. The model shows that leader type acts mainly through how often drivers close up behind a truck rather than through how they follow once they have, and that drivers differ from one another far more than their following differs between leader types.

The paper is organised as follows. Section 2 reviews related work. Section 3 describes the experiment, measures and analysis. Section 4 presents the results, Section 5 discusses their implications and limitations, and Section 6 concludes.

## 2 Literature review

### 2.1 AVDL configuration and ramp influence areas

AVDL performance depends on lane extent, access arrangement, network location and traffic composition. Rad et al. (2020) identified minimum dedicated-lane length and access design as conditions for stable automated operation. Network-level studies have formulated dedicated-lane deployment as an optimisation problem under fixed or elastic demand (Madadi et al., 2021; Zhang et al., 2023; Ngoduy et al., 2024), and capacity analyses show that the benefit of lane management depends on penetration (Ghiasi et al., 2017; Chen et al., 2017). Dedicated lanes tend to become advantageous only as AV penetration increases (Papadimitriou et al., 2022; Chang and Chen, 2025).

Ramp influence areas add a spatial dimension. Placing a dedicated lane on the offside lane minimises reconstruction but forces vehicles entering from or leaving to the ramps to cross the general-purpose lanes (Vander Laan and Sadabadi, 2017; Kumar et al., 2020). He et al. (2022) found limited merging-zone safety improvement at low AV penetration, and Eom and Park (2025) showed that higher penetration does not necessarily improve weaving efficiency when automated vehicles accept large gaps. Liu et al. (2025) optimised AVDL entry and exit locations and buffer-zone lengths in ramp influence areas. All of these studies are model-based, and the human-driven vehicles in them follow car-following and lane-changing rules whose parameters do not depend on the layout or on the vehicles around them.

## 2.2 Human drivers near automated trucks and platoons

Heavy vehicles affect capacity and the behaviour of surrounding drivers through their size and dynamics (Al-Kaisy et al., 2002; Peeta et al., 2005). Lead-vehicle type also affects following behaviour. In naturalistic and instrumented-vehicle data, drivers followed light trucks, vans and trucks at shorter distances or time headways than passenger cars, which has been attributed to the reduced view of traffic ahead (Sayer et al., 2000; Brackstone et al., 2009). Truck-platooning research has focused mainly on within-platoon control and operator workload (Nilsson et al., 2017; Neto et al., 2024), and reviews have identified the responses of surrounding light-vehicle drivers as an unresolved issue (Zhao et al., 2024). Driving-simulator studies show that manual drivers adapt their headways when driving next to platoons with short internal gaps (Gouy et al., 2014) and alter their car-following near dedicated lanes (Schoenmakers et al., 2021; Razmi Rad et al., 2024). These studies typically compare the presence and absence of automated vehicles rather than varying their share, and they rarely separate behavioural adaptation from the composition of the traffic that drivers follow.

## 2.3 Car-following models estimated from driving-simulator data

Car-following models have been estimated from three kinds of data, which trade realism against control. Naturalistic driving data record everyday behaviour; Sangster et al. (2013), for example, calibrated four models on more than 2,000 car-following events from eight drivers on a multilane highway. Experiments with instrumented vehicles driving as a platoon on real roads provide repeated observations under controlled conditions, and in high-speed platoon runs the same driver behaved quite differently from one run to the next (Huang et al., 2018).

Driving simulators give the most control. They allow car-following to be observed under conditions that are rare, unsafe or unobservable in field trajectories, and they allow model parameters to be linked to driver states. Hoogendoorn et al. (2011) estimated psycho-spacing models from simulator drives in fog and found shifts in drivers' action points, together with large differences between drivers. Paschalidis et al. (2019) combined simulator and physiological data and showed that stress changes how drivers accelerate and brake behind a leader. In mixed traffic, simulator studies show that drivers shorten their time headways next to automated platoons and dedicated lanes (Gouy et al., 2014; Schoenmakers et al., 2021) and as the share of simulated automated vehicles rises (de Zwart et al., 2023). Reddy et al. (2026) calibrated the original Intelligent Driver Model (IDM), a car-following model proposed by Treiber et al. (2000), and an extended IDM in a driving simulator. The authors suggested that drivers following an AV adopted shorter safe time headways and smaller jam spacings than when following a human-driven vehicle.

Empirical trajectory studies show that car-following behaviour differs systematically between drivers and that a single parameter set cannot represent a driver population (Ossen et al., 2006; Ossen and Hoogendoorn, 2011; Saifuzzaman and Zheng, 2014). Mixed-traffic analyses of AV impacts nonetheless commonly represent human-driven vehicles with one calibrated model (Talebpour and Mahmassani, 2016; Chen et al., 2017). Simulator experiments with repeated measures can estimate how much variation is between drivers and how much is within drivers across conditions, but such estimates are seldom reported for experiments involving automated vehicles.

Field trajectories also show that following behaviour depends on the type of the vehicle ahead (Aghabayk et al., 2016). To our knowledge, however, no simulator-based model has distinguished automated trucks from cars as leaders.

Simulator-based models carry well-known caveats. First caveats relate to validity, as behavioural validity can be absolute, when simulator and road values agree, or relative, when differences between conditions agree in direction and approximate size (Kaptein et al., 1996; Godley et al., 2002). The relationship between simulator fidelity and validity was not straightforward, so a more sophisticated simulator does not guarantee more valid behaviour (Wynne et al., 2019).

The second caveat relates to transferability. Risto and Martens (2014) found no significant difference between the headways drivers chose in a simulator and on the road. Car-following models estimated in simulators, however, transfer to field traffic in their structure rather than in their parameter values (Papadimitriou and Choudhury, 2017). Paschalidis et al. (2021) tested methods for improving transferability. Compared with field trajectories, simulator drivers decelerated less, responded more slowly and were less sensitive to closing speed, and transfer required combining simulator and field data (see also Hoogendoorn and Hoogendoorn, 2010). Both studies argue for caution in applying simulator-derived parameters directly to field traffic.

The third caveat relates to distance perception. Renner et al., (2013) reported that in head-mounted displays, egocentric distances in virtual environments are typically underestimated, on average, to about three-quarters of the modelled distance.

The fourth caveat relates to the drives' duration. Simulator drives are short, and leaders often travel at steady speeds, so not all parameters of a model such as the IDM can be identified. Sensitivity analysis shows that parameters to which the fit is insensitive can be fixed without loss (Punzo et al., 2015). Estimates also depend on the data and the calibration set-up. Which driving regimes the trajectories cover affects calibration and validation, and for the IDM the standstill regime matters (Sharma et al., 2019). Measurement errors can bias estimates and make the objective function less sensitive, so the parameters that minimise it need not reproduce the following dynamics best (Ossen and Hoogendoorn, 2008). Results also depend on the error measure and the simulation set-up (Punzo et al., 2021). Treiber and Kesting (2013) examined how estimates respond to the sampling rate, measurement errors and smoothing, found gap-based error measures more robust than speed-based ones, and showed that parameters can be correlated. Combining individual estimates into a representative parameter set for a population of drivers therefore requires care, and methods of doing so perform differently (James et al., 2019).

The fifth caveat relates to model calibration. Models calibrated on routine following may not hold in safety-critical situations. Berghaus et al. (2021) calibrated four car-following models for each driver on simulator data. In an emergency-braking scenario in which drivers were momentarily shocked, all four models underestimated the drivers' deceleration.

Simulator-calibrated parameters are therefore most defensible for relative comparisons, such as between leader types or between drivers, rather than as absolute values for field simulation. A close fit to the calibration trajectories, or to held-out trajectories from the same experiment, shows how well a model performs within that experiment.

### 2.5 Research gaps

Four gaps follow from this review: (i) No human-participant evidence is available on how alternative AVDL entry and exit arrangements relative to ramp weaving sections affect human driving behaviour; (ii) Effects of automated-truck penetration on surrounding drivers have not been decomposed into behavioural adaptation; (iii) Between-driver heterogeneity in responses to AVDL designs and automated trucks has not been quantified; (iv) Simulator-based car-following models have distinguished automated from human-driven cars as leaders (Reddy et al., 2026), but not automated trucks from cars, so the models used to evaluate AVDLs cannot represent the leader-composition mechanism.

## 3 Methodology

### 3.1 Participants and assignment

This study recruited licensed drivers through email invitations to university students. Exclusion criteria were a history of epilepsy, motion sickness, vertigo or nausea; participation in a driving-simulator study in the previous six months; and uncorrected visual impairment. The study was approved by the University of New South Wales Human Research Ethics Advisory Panel (approval iRECS8743), and all participants gave written informed consent. Sixty-three drivers (30 women, 33 men) aged 18–45 years took part in the study. Participants were aged between 18 to 45 (25 participants were aged 18–23 years, 26 were 24–29, 9 were 30–35, 2 were 36–41 and 1 was 42–45).

AVDL layout was a between-participants factor and automated-truck penetration a within-participants factor. Participants were randomly assigned to one of the four layouts, and the order of the four penetration levels was randomised for each participant. Testing of all four layouts was interleaved over the same two-week period (23 October to 5 November 2025). All 63 participants completed at least one valid drive (16, 17, 15 and 15 in layouts a–d; Table 1). The layouts did not differ significantly in gender (Fisher's exact test, $p = 0.59$) or age band (18–23, 24–29 and 30–45 years; Fisher's exact test with $10^5$ Monte Carlo samples, $p = 0.33$). With 15–17 participants per group, layout (a) nonetheless included proportionally more women (10 of 16) and more participants aged 30 or over (6 of 16). Gender and age were therefore added as covariates in a sensitivity analysis in Section 3.4. In total, 241 full drives were retained for the analysis.

**Table 1. Participants, demographics, drives and data availability by AVDL layout.**

| AVDL layout and length | Drivers | Women/men | Age 18–23/24–29/30–45 years | Drives |
|---|---|---|---|---|
| (a) 1800 m | 16 | 10 / 6 | 5 / 5 / 6 | 63 |
| (b) 1500 m | 17 | 7 / 10 | 6 / 10 / 1 | 62 |
| (c) 1200 m | 15 | 7 / 8 | 6 / 6 / 3 | 56 |
| (d) No AVDL | 15 | 6 / 9 | 8 / 5 / 2 | 60 |
| Total | 63 | 30 / 33 | 25 / 26 / 12 | 241 |

## 3.2 Simulation Apparatus and Scenario Design

The simulator comprised a fixed-base driving seat with a steering wheel and pedals and a head-mounted virtual-reality display with integrated eye tracking. The headset has an HTC Vive Pro Eye operating at 90 Hz. The headset is equipped with an eye tracker SRanipal SDK that was calibrated before each driving scenario. Each frame, the simulator cast the right-eye gaze ray into the scene and recorded the name of, and hit position on, the first object it intersected. Vehicle state, the forward range to the nearest vehicle ahead (ray-cast up to 200 m), the simulator's time-to-collision value and the gaze hit were written to a log every 0.1 s (10 Hz). Heart rate was recorded with a Polar H10 chest strap sensor and exported from Polar Flow at 1 Hz. The virtual driving environment was built in Unity.

After giving consent and fitting the heart-rate sensor, participants completed a short familiarisation drive and the eye-tracker calibration. They then completed four drives, one at each penetration level, in a randomised order. Serial position is defined as the order a drive came in the participant's session (i.e. first, second, third or fourth). The serial position matters because the penetration levels were given in random order. One participant might drive the 80% condition first, and another might drive it last. Drivers also change over a session as they get used to the simulator, may tire, or they may become faster and keep shorter mean gaps from one drive to the next. Because orders were randomised, each penetration level appeared in each serial position between 8 and 22 times across the analysed drives, so serial position was included as a covariate in all models. Each drive took about 8 min (median 493 s). Participants were instructed to drive as they normally would and to use the general-purpose lanes. After each drive, they completed the Simulator Sickness Questionnaire (SSQ) and the NASA Task Load Index (NASA-TLX). A session lasted about 65 min. The interval between the end of one drive and the start of the next, which included the questionnaires and any restart, had a median of 4 min (interquartile range 2.6–6.0 min), and the four drives spanned a median of 46 min.

The simulated motorway had three general lanes with left-hand traffic and was 12.7 km long. It comprised a 2-km warm-up section followed by three 1.5-km weaving sections (2.0–3.5, 6.5–8.0 and 11.0–12.5 km), each bounded by an on-ramp merge and an off-ramp diverge and separated by 3-km recovery sections (Figure 1). In layouts a–c, the offside (rightmost) lane was designated as an AVDL in each weaving area, and only automated vehicles were permitted to use it. The layouts differed only in where the AVDL started and ended relative to the weaving section (Figure 2). In layout (a) (1800 m), the AVDL extended 150 m upstream of the merge and 150 m downstream of the diverge, so its entry and exit lay outside the weaving section and the dedicated lane overlapped the whole ramp influence area. In layout (b) (1500 m), the AVDL coincided with the weaving section. In layout (c) (1200 m), the AVDL began 150 m downstream of the merge and ended 150 m upstream of the diverge, offsetting its entry and exit from the ramp gores. The layout (d) has no AVDL.

Short entry and exit transition zones (100 m, posted 80 km/h) marked where eligible vehicles should prepare to enter or leave the AVDL; the AVDL itself was posted at 100 km/h. Overhead signs indicated the lane restriction. Weather and lighting were fixed at clear daytime conditions.

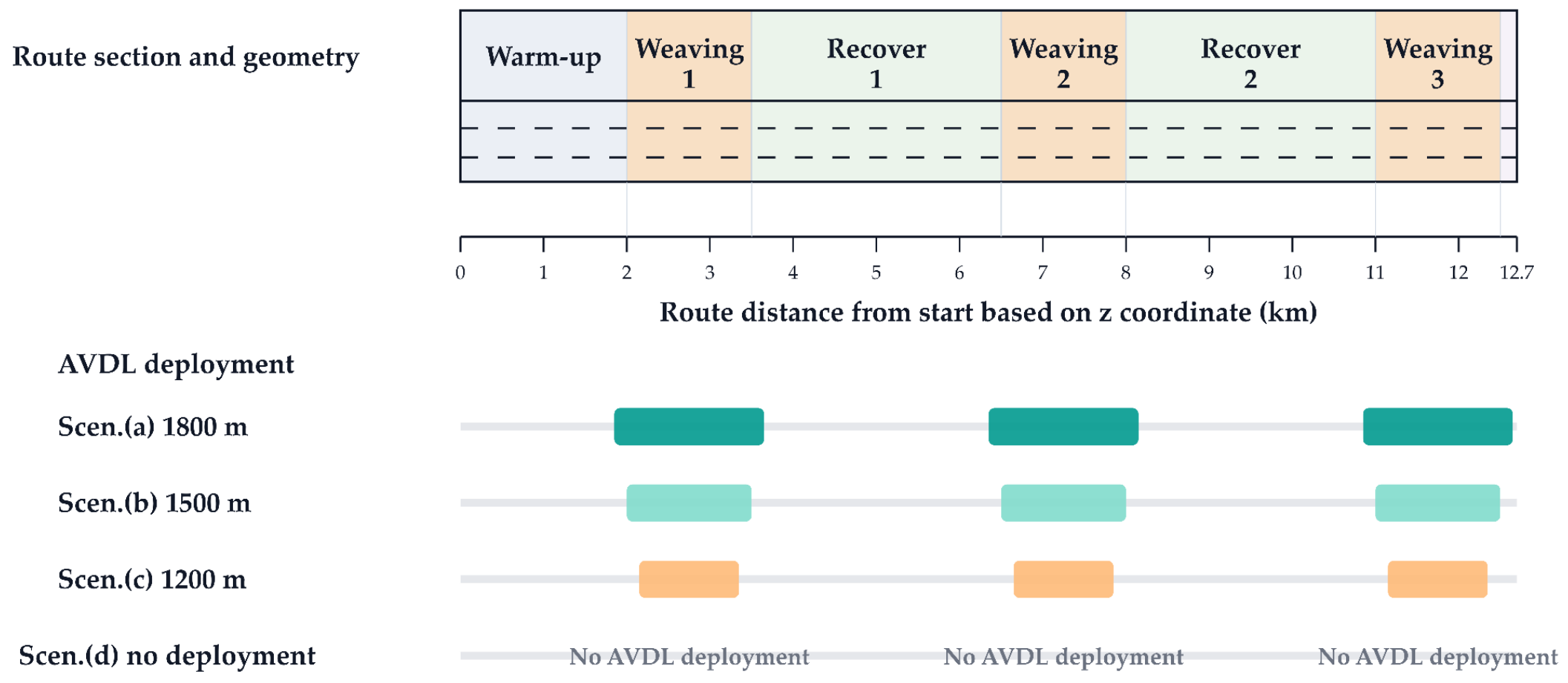

**Figure 1. Route geometry and AVDL deployment layouts.** Route distance is measured from the start of the simulated motorway. The three weaving sections each begin at an on-ramp merge and end at an off-ramp diverge.

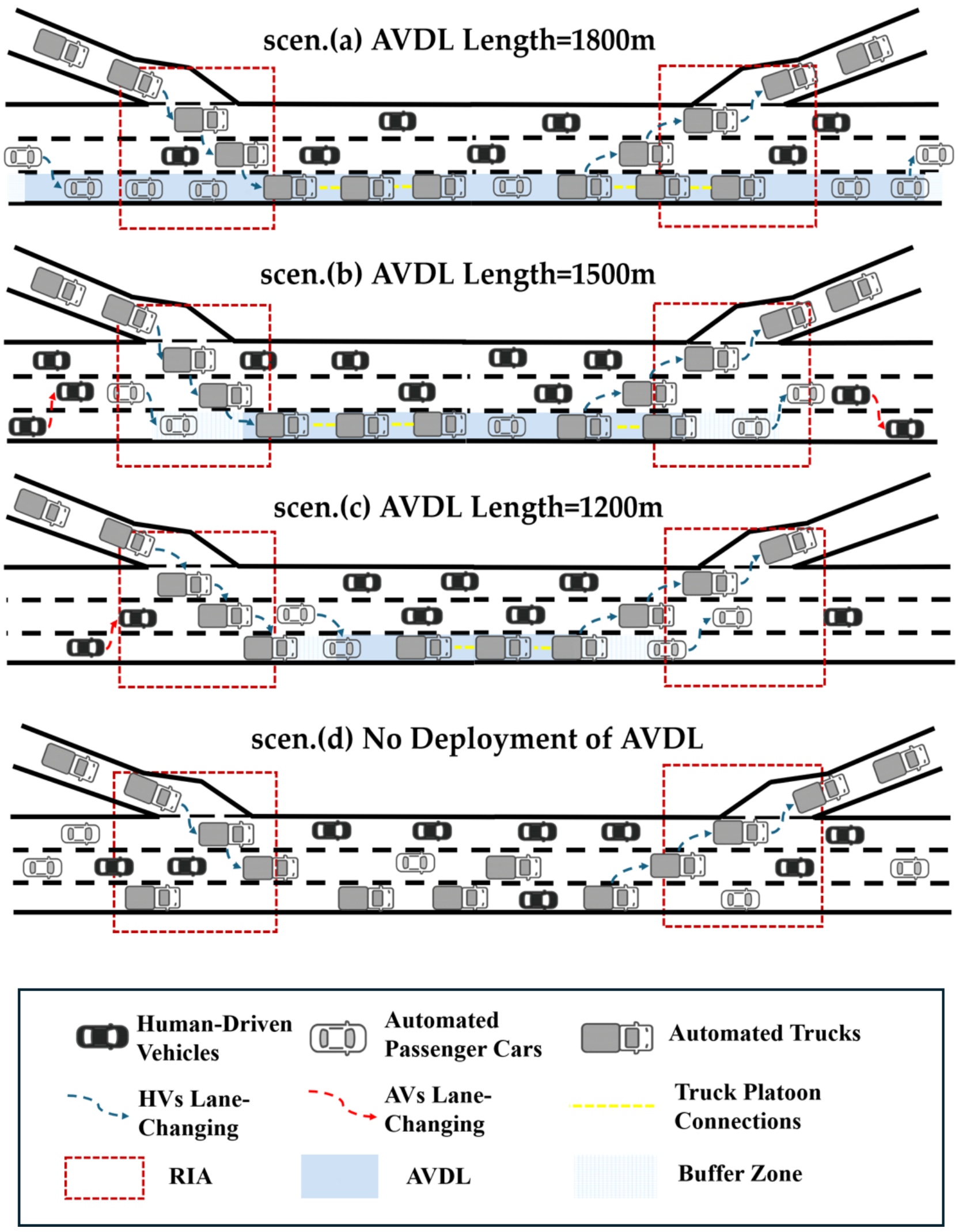

**Figure 2. Schematic of the four AVDL layouts relative to the ramp weaving areas.**

Background traffic was generated stochastically. At each vehicle-generation event, the new vehicle was an automated truck with probability $p$, where $p \in \{0.20, 0.40, 0.60, 0.80\}$ defined the penetration condition, and otherwise another background vehicle. The realised share of automated trucks therefore varied around $p$ within a drive, and was implemented by independent Bernoulli sampling at vehicle generation. If $n$ vehicles were generated in a drive, the number of automated trucks was $N_{AT} = \sum_{i=1}^{n} X_i \sim \text{Binomial}(n, p)$, and the realised share $\hat{p} = N_{AT}/n$ varied around $p$ from drive to drive. The simulator build implements car longitudinal control through a Gipps car-following state with PID speed tracking.

Automated trucks were 15 m long, travelled in groups of two to three vehicles and used the AVDL where one was present. After entering the AVDL, automated trucks were governed by the cooperative adaptive cruise control (CACC) model proposed by Liu et al. (2024), and could travel cooperatively as truck platoons. Platooning was therefore an operational state resulting from the longitudinal control strategy on the AVDL, rather than the unit used in the vehicle-generation process. Once automated platooned trucks approach the AVDL, they must merge into the AVDL. When the vehicle approaches the off-ramp area, the AV can choose to exit or continue to drive on the highway.

The CACC-based longitudinal model (eq. 1-4) determines the follower truck's acceleration and updated speed based on the spacing error, relative speed, and desired time headway with respect to the preceding truck. Through this control process, automated trucks can maintain stable cooperative following and form operational platoons while travelling on the dedicated lane.

$$a(t) = k_1\left[v_{follow}(t)h_{truck}(t) - h_{desired}v_{follow}(t)\right] + k_2\left[v_{lead} - v_{follow}\right] \quad (1)$$

$$h_{truck}(t) = \frac{x_{lead}(t) - x_{follow}(t) + L_{truck}}{v_{follow}(t)} \quad (2)$$

$$e = x_{lead}(t) - x_{follow}(t) - h_{desired}v_{follow}(t) \quad (3)$$

$$v(t + \Delta t) = v_{follow} + k_p e + k_d \dot{e} \quad (4)$$

where $k_1$ and $k_2$ are control gain for truck platoon acceleration, calibrated by Liu et al. (2024), valued at -0.23$s^{-1}$ and 0.07$s^{-1}$. The parameter $e$ is the spacing error between the actual spacing and the desired spacing, and $\dot{e}$ is the derivative of $e$ with respect to time. $k_p$ and $k_d$ are control gain for truck platoon speed, calibrated by Liu et al. (2024), valued at 0.45$s^{-1}$ and 0.25$s^{-1}$. $x_{lead}(t)$ and $x_{follow}(t)$ are the real-time longitudinal positions (m) of the leader and follower truck. Similarly, $v_{lead}(t)$ and $v_{follow}(t)$ are the instantaneous speeds of the leader and follower truck. $h_{truck}$ is the actual headway, and $h_{desired}$ is the desired time headway between truck platoons, assuming 3s (Gouy et al., 2014). $a(t)$ is the real-time acceleration and $L_{truck}$ is the length of the truck, assuming 15 meters. $\Delta t$ is the simulation time step (0.1s). Once automated platooned trucks approach the AVDL, they must merge into the AVDL. When the vehicle approaches the off-ramp area, the AV can choose to exit or continue to drive on the highway.

### 3.3 Simulation Outcome Measures

A number of outcome measures and indicators were analysed in this study to address the research questions. Table 2 outlines the simulation outcome measures and their roles in the analysis. Multiplicity was controlled within pre-defined families of outcomes (last column of Table 2), because testing many outcomes raises the chance that some effects appear significant by chance alone. For each model term, two procedures were used, matched to the role of the family. Holm's procedure (Holm, 1979) controlled the family-wise error rate, the probability of any false positive, across the three primary outcomes. The Benjamini–Hochberg (BH) procedure (Benjamini and Hochberg, 1995) controlled the false discovery rate, the expected share of false positives among significant results, within each secondary or exploratory family.

**Table 2. Outcome measures, definitions and analysis roles.**

| Domain | Measure | Definition | Role (multiplicity family) |
|---|---|---|---|
| Safety | Short time-gap exposure | Share of following samples with time gap < 1.5 s (Eq. 6) | Primary (Holm across the three primary outcomes) |
| Safety | Gap < 1.0 s | As above, with a 1.0-s threshold | Secondary (BH, safety) |
| Safety | Mean time gap | Mean of Eq. (5) over following samples | Secondary (BH, safety) |
| Safety | TTC exposure | Share of samples with 0 < TTC < 3 s (Eq. 7) | Secondary (BH, safety) |
| Safety | Rear-end conflict rate | Events with 0 < TTC ≤ 1.5 s per km driven, excluding 300-m pre-merge zones | Secondary (BH, safety) |
| Efficiency | Weaving-section travel time | Mean time to traverse the three 1.5-km weaving sections | Primary |
| Efficiency | Full-route travel time | Time from 0 to 12.6 km | Secondary (BH, efficiency) |
| Efficiency | Weaving-section speed | Mean speed within the weaving sections | Secondary (BH, efficiency) |

| Domain | Measure | Definition | Role (multiplicity family) |
|---|---|---|---|
| Workload | Raw NASA-TLX | Mean of six subscales (performance reversed) × 10 | Primary |
| Comfort | SSQ total severity | Kennedy et al. (1993) weights; log(1 + x) in models | Secondary |
| Mechanism | Truck-following share; gap < 1.5 s by leader type | The share of following time behind a truck | Exploratory |
| Workload | Six NASA-TLX subscales | 0–10 | Exploratory (BH, subscales) |
| Physiology | Mean HR; within-drive SD of HR | 1-Hz samples, whole drive | Exploratory (BH) |
| Gaze | AOI dwell (cars, trucks, road and roadside, unknown targets); truck share of vehicle gaze; vehicle visits; AOI switching; stationary and transition entropy; horizontal gaze dispersion | Drives with ≥ 50% gaze hits; AOI mapping in Appendix E | Exploratory (BH, gaze) |

**Car-following safety margins:** A sample was classified as *following* when a vehicle was detected ahead within 200 m and at a speed of over 5m/s. The time gap is calculated as $\tau_t = g_t/v_t$, where $g_t$ is the forward range from the front of the participant's vehicle to the rear of the vehicle ahead in the same lane at sample $t$, and $v_t$ the participant's speed. Short time-gap exposure ($E_{1.5}$), the primary safety outcome, is considered as the share of following samples with $\tau_t < 1.5$ s. The 1-second threshold and the mean time gap over following samples were analysed as secondary outcomes. Time-to-collision with the vehicle ahead was computed by the simulator as:

$$TTC_t = \frac{g_t}{v_t - v_{lead,t}}, v_t > v_{lead,t} \tag{5}$$

TTC exposure was summarised as the share of all samples in a drive with $0 < \mathrm{TTC}_t < 3$ s, a normalised form of time-exposed TTC (Minderhoud and Bovy, 2001).

**Leader type:** For each drive, we computed the share of following time behind a truck and $E_{1.5}$ separately for car-following and truck-following samples. Drives with fewer than 50 samples of a leader type were excluded from that leader type's estimate.

**Rear-end conflicts:** The Surrogate Safety Assessment Model (SSAM; Gettman and Head, 2003; Gettman et al., 2008) identifies conflicts between all pairs of vehicles from simulated trajectories. The simulator logged only the participant's vehicle, so SSAM itself could not be applied. Instead, its default rear-end criterion (TTC ≤ 1.5 s) was applied to the participant and the vehicle directly ahead. A conflict was a run of consecutive samples with $0 < \mathrm{TTC}_t \leq 1.5$ s while the participant travelled at 2 m/s or more. A jump of more than 5 m in forward range, which indicates a new leader, started a new event. For each event, we recorded the minimum TTC, the maximum closing speed ($g_t/\mathrm{TTC}_t$), the participant's speed reduction from 0.5 s before to 1.5 s after the event, the leader type and whether the simulator logged a contact during the event. The conflict rate, in events per kilometre driven, was a secondary safety outcome. Events beginning within 300 m upstream of a merge were excluded, and the same 300 m was removed from the distance used as exposure. Because these excluded zones contributed 12.4% of car-following samples, short time-gap exposure was recomputed without them as a sensitivity analysis.

**Traffic efficiency:** Travel time was measured over road sections of identical length in all layouts: the mean time to traverse each of the three 1.5-km weaving sections (primary), the full-route travel time, and the mean speed within the weaving sections. Travel time over each layout's own AVDL segment was not used, because those segments differ in length (1200–1800 m).

**Workload and simulator sickness.** The NASA-TLX (Hart and Staveland, 1988) was administered with its six subscales (mental, physical and temporal demand, performance, effort and frustration), each rated on a 0–10 scale. Pairwise weighting was not administered, so an unweighted (raw) TLX score was computed (Hart, 2006). The performance item asked how successful participants had been, so it was reverse-scored before averaging, and the mean of the six subscales was multiplied by 10 to give a 0–100 score. The 16 SSQ symptoms were rated None, Slight, Moderate or Severe and coded 0–3, and the nausea, oculomotor and disorientation subscales and the total severity score were computed with the weights of Kennedy et al. (1993).

**Heart rate:** For each drive, mean heart rate and its within-drive standard deviation were computed from 1-Hz samples between 40 and 200 bpm over the whole recording (Shaffer and Ginsberg, 2017).

**Gaze:** The gaze ray had unlimited range and stopped at the first collider it met, including invisible trigger volumes used to control traffic. Samples were therefore classified into four areas of interest (AOIs): (i) cars; (ii) trucks; (iii) road and roadside, comprising the terrain (which includes the road surface), trees, barriers, poles and signs; (iv) unknown targets (the invisible walls and boxes).

Samples in which the ray hit nothing were treated as missing. For each drive, the following measures were computed: (i) the dwell share of each AOI among samples with a hit, and trucks' share of vehicle gaze; (ii) vehicle gaze visits per minute and their mean duration, where a visit is a run of consecutive samples on one AOI; (iii) AOI switches per minute; (iv) stationary and transition gaze entropy, normalised to 0–1 (Krejtz et al., 2015; Shiferaw et al., 2019); (v) three measures of the dispersion of the horizontal gaze angle relative to the route direction: normalised entropy over 2° bins, the share of samples in the densest 5% of bins, and the standard deviation.

The gaze direction is valid even when the ray met an invisible object, so the angle measures use all hits.

The gaze log was written at 10 Hz and recorded no fixations, so visits are runs of samples rather than fixations. Gaze outcomes were analysed for drives in which at least half of the samples hit an object, and all drives were used in a sensitivity analysis. To interpret the logged object names, we inspected the scene files of the simulator build (Appendix, Table A2).

For a drive with dwell shares $p_i$ on the four AOIs among samples with a hit, the normalised stationary gaze entropy is calculated as:

$$H_s = -\frac{1}{\log_2 4}\sum_{i=1}^{4} p_i \log_2 p_i \tag{6}$$

Collapsing consecutive samples on the same AOI gives the sequence of AOI visits. With $n_{ij}$ the number of transitions from AOI $i$ to AOI $j \neq i$, $p_{ij} = n_{ij}/\sum_j n_{ij}$ and $\pi_i = \sum_j n_{ij} / \sum_{i,j} n_{ij}$, the normalised transition entropy is calculated as:

$$H_t = -\frac{1}{\log_2 3}\sum_{i} \pi_i \sum_{j\neq i} p_{ij} \log_2 p_{ij} \tag{7}$$

Both measures range from 0, when gaze stays on one AOI or follows one transition pattern, to 1, when it is spread evenly (Krejtz et al., 2015; Shiferaw et al., 2019). For visit rate and duration, a visit was a run of consecutive samples on one AOI, ended by a change of AOI or by a sample without a hit.

## 3.4 Statistical analysis

Each outcome was analysed with a linear mixed-effects model with fixed effects for layout (four levels), penetration (four levels), their interaction and the serial position of the drive (linear), and a random intercept for participant:

$$y_{ij} = \mu + \alpha_{L(i)} + \beta_j + (\alpha\beta)_{L(i)j} + \gamma\, o_{ij} + u_i + \varepsilon_{ij}, \quad u_i \sim N(0, \sigma_u^2),\ \varepsilon_{ij} \sim N(0, \sigma_\varepsilon^2) \tag{8}$$

where $i$ indexes participants, $j$ penetration levels, $L(i)$ is participant $i$'s layout and $o_{ij}$ the serial position. The random intercept models the repeated measurements on each participant, so the layout effect is tested against between-participant variation and the penetration and interaction effects against within-participant variation. Models were fitted by restricted maximum likelihood in R with lme4 and lmerTest (Bates et al., 2015; Kuznetsova et al., 2017), using sum-to-zero contrasts and Type III F-tests with Kenward–Roger degrees of freedom (Kenward and Roger, 1997). The random intercept implies a compound-symmetric covariance among each participant's four drives. Sphericity corrections apply to repeated-measures ANOVA and are not used in this framework, so departures from compound symmetry were examined instead by refitting the models with a random slope for penetration (see below). The same random slope was used to test whether drivers differed in how they responded to penetration. Its variance was tested with a likelihood-ratio test on maximum-likelihood fits, with the p-value halved because a variance is tested at its boundary of zero, and adjusted within families as for the fixed effects. Residuals of the bounded outcomes departed from normality, so the primary safety outcome was also analysed on the logit scale. Partial $\eta^2$ was computed from the F-statistics, and the share of variance attributable to fixed effects and to participants was computed following Nakagawa and Schielzeth (2013). Post-hoc comparisons of layouts used Holm-adjusted pairwise contrasts of estimated marginal means, and penetration trends were tested with a linear polynomial contrast. Conflict counts per drive were analysed with a negative-binomial generalised linear mixed model, with a log link and quadratic variance (NB2), fitted by maximum likelihood, integrating over the participant random effects with the Laplace approximation implemented in glmmTMB R package (Brooks et al., 2017). The model had the same fixed effects and random intercept as Eq. (4), with the log of the distance driven as an offset.

Many layout × penetration cells contained few events, so layout and penetration were tested with Wald $\chi^2$ tests in the model without the interaction, and the interaction with a likelihood-ratio test. Rates are reported as estimated marginal means per 100 km. Two further definitions were analysed in the same way and are reported descriptively: conflicts within the weaving sections (offset: weaving-section distance) and severe conflicts (minimum TTC ≤ 1.0 s).

Multiplicity was controlled within pre-defined families. For each model term, Holm's procedure (Holm, 1979) was applied across the three primary outcomes, and the Benjamini–Hochberg false-discovery-rate procedure (Benjamini and Hochberg, 1995) within each family of secondary or exploratory outcomes (Table 2). Exploratory results are reported as such and are not used for confirmatory claims. Proportions were analysed on their original scale, and SSQ scores on the log(1 + x) scale. Sensitivity analyses refitted the models with a random slope for linear penetration (Barr et al., 2013; Matuschek et al., 2017), without the serial-position covariate, with gender and age (band midpoint) as participant-level covariates, with logit-transformed short-gap exposure, and on complete drives only.

The planning calculation (G*Power; Faul et al., 2007) assumed f = 0.25, α = .05, power = .80, four groups, four drives and a correlation of 0.5 among repeated drives, and gave a required total of 116 participants for the between-participants (layout) effect. The achieved sample of 63 is about half of that. A sensitivity analysis under the same assumptions shows that, with 58–63 participants, effects of f ≥ 0.35 (between-participants), f ≥ 0.15 (within-participants) and f ≥ 0.19 (interaction) could be detected with 80% power, rising to f ≥ 0.40 for the layout effect if the correlation among repeated measures is 0.7. The layout comparisons can therefore detect only large effects, and their null results are interpreted through confidence intervals rather than as evidence of no effect (Lakens, 2022).

### 3.5 Leader-type-dependent car-following model

To express the leader-type difference in a form that can be used in traffic simulation, we calibrated the IDM (Treiber et al., 2000) on close-following episodes. A car-following episode is a continuous stretch of at least 15 s within a drive in which the participant followed the same leader in the same lane; a drive can contain several episodes or none. An episode ended at a jump of more than 3 m in forward range, and episodes with more than 1.5 m of lateral movement were excluded. Position, speed and forward range were logged every 0.1 s. The leader's speed was not logged. It was obtained as the participant's speed plus the rate of change of the gap, after both signals had been smoothed with a second-order Savitzky–Golay filter over 1.1 s, and it agreed with the closing speed implied by the logged TTC to within 0.12 m/s (median absolute difference). The calibration error was computed on the unsmoothed gap, and the model was integrated with the logging step of 0.1 s. The IDM acceleration is

$$\frac{dv}{dt} = a\left[1 - \left(\frac{v}{v_0}\right)^4 - \left(\frac{s^*(v,\Delta v)}{s}\right)^2\right], \qquad s^*(v,\Delta v) = s_0 + \max\left(0,\ vT + \frac{v\,\Delta v}{2\sqrt{ab}}\right) \tag{9}$$

where $s$ is the gap, $v$ the participant's speed and $\Delta v$ the rate at which the participant approaches the leader. The desired speed $v_0$, maximum acceleration $a$, comfortable deceleration $b$ and standstill gap $s_0$ were common to all drivers, because estimating every parameter for every driver is neither necessary nor reliable (Punzo et al., 2015). The desired time gap $T$ was specified in four ways: (i) one value for all drivers (population model); (ii) one value per driver (driver model); (iii) one value per driver, multiplied by a common factor $\theta$ when the leader was a truck (leader-type model); and (iv) separate values for each driver and leader type.

Speeds in close following varied little, therefore $s_0$ and $T$ could not be estimated separately for each leader type, and leader type was allowed to act through $T$ only.

Each episode was simulated from its observed initial state against the observed leader trajectory. The parameters minimised the normalised root-mean-square error (NRMSE) of the gap (Punzo et al., 2021), with the driver-level and common parameters estimated in turn until the error changed by less than $10^{-5}$. The specifications were compared on held-out data: alternate episodes of each driver and leader type were used for calibration, and the rest were held out. Both sets come from the same experiment, so this comparison tests prediction within the simulator experiment, not validity in real traffic.

Because parameter estimates can depend on the error measure, the processing of the trajectories and the model form (Ossen and Hoogendoorn, 2008; Treiber and Kesting, 2013), the calibration and the episode-level analysis below were repeated with alternative choices (Appendix, Table A3). The alternative objectives were the speed NRMSE and an equally weighted combination of spacing and speed errors. The alternative processing used smoothing windows of 0.5 s and 2.1 s. The alternative selections kept only episodes with a median time gap below 2 s, episodes of at least 30 s, or episodes away from the pre-merge zones. The alternative model form was an IDM whose acceleration responds to the state 0.5 s or 1.0 s earlier.

The leader-type effect was also estimated episode by episode. $T$ was fitted to each episode using the common parameters of the driver model, and $\log T$ was analysed with a linear mixed model with leader type, penetration, drive order and layout as fixed effects and random intercepts for driver and for drive within driver, because episodes from the same drive are not independent. The variance of $\log T$ not explained by the fixed effects was split between drivers, between drives of the same driver and between episodes within a drive. Whether drivers differed in their response to truck leaders was tested with a random slope for leader type, as for penetration in Section 3.4, and the episode's median speed was added as a situational covariate. Finally, the share of following time at a time gap below 3 seconds was computed for each drive and leader type and compared between leader types with a linear mixed model with random intercepts for driver and drive.

# 4 Results

## 4.1 Car-following safety margins

Table 3 summarises the tests for the primary and secondary outcomes. The results suggest that automated-truck penetration had a significant effect on short time-gap exposure ($p = .002$; Holm-adjusted $p = .006$). Exposure fell monotonically from 28.5% of following time at 20% penetration to 18.1% at 80% (linear trend $p < .001$; Table 4, Figure 3A). The result was unchanged with a random slope for penetration ($p = .003$), without the serial-position covariate ($p = .002$), with gender and age as covariates ($p = .002$), with a logit transformation ($p = .001$), on complete drives only ($p = .006$) and without the following samples in the 300-m pre-merge zones ($p = .004$; 27.9% to 17.8%).

The share of following time below 1.0 s showed the same pattern (13.9% to 7.7%; BH-adjusted $p = .007$), and the mean time gap rose from 3.25 s to 3.59 s, although this change was not significant after correction (BH-adjusted $p = .31$).

AVDL layout had no detectable effect on the time-gap or TTC measures, and the layout × penetration interaction was not significant for short-gap exposure. TTC exposure was low in all conditions (about 1% of drive time) and was not affected by layout or penetration after correction.

**Table 3. Mixed-model tests for primary and secondary outcomes.**

| Outcome | Layout | Penetration | Layout × penetration |
|---|---|---|---|
| *Primary outcomes* | | | |
| Short time-gap exposure (gap < 1.5 s) | $F(3, 59) = 0.03$; $p = 0.992$ (1.000) | $F(3, 167) = 5.13$; $p = 0.002$ (0.006) | $F(9, 166) = 0.85$; $p = 0.575$ (1.000) |
| Travel time per 1.5-km weaving section (s) | $F(3, 59) = 0.29$; $p = 0.836$ (1.000) | $F(3, 167) = 1.23$; $p = 0.302$ (0.604) | $F(9, 167) = 0.90$; $p = 0.524$ (1.000) |
| Raw NASA-TLX (0–100) | $F(3, 54) = 2.36$; $p = 0.081$ (0.244) | $F(3, 155) = 0.35$; $p = 0.786$ (0.786) | $F(9, 155) = 1.12$; $p = 0.353$ (1.000) |
| *Secondary outcomes* | | | |
| Mean time gap when following (s) | $F(3, 59) = 0.03$; $p = 0.994$ (0.994) | $F(3, 166) = 1.78$; $p = 0.154$ (0.307) | $F(9, 166) = 0.32$; $p = 0.968$ (0.968) |
| Share of drive time with TTC < 3 s | $F(3, 58) = 0.27$; $p = 0.845$ (0.994) | $F(3, 168) = 0.39$; $p = 0.757$ (0.976) | $F(9, 168) = 1.97$; $p = 0.045$ (0.181) |
| Gap < 1.0 s share of following time | $F(3, 59) = 0.04$; $p = 0.990$ (0.994) | $F(3, 167) = 5.27$; $p = 0.002$ (0.007) | $F(9, 167) = 1.13$; $p = 0.343$ (0.687) |
| Rear-end conflict rate (TTC ≤ 1.5 s; events per km) | $\chi^2(3) = 6.98$; $p = 0.073$ (0.290) | $\chi^2(3) = 0.21$; $p = 0.976$ (0.976) | $\chi^2(9) = 6.74$; $p = .664$ (.885) |
| Full-route travel time (s) | $F(3, 58) = 0.63$; $p = 0.599$ (0.875) | $F(3, 161) = 0.15$; $p = 0.929$ (0.929) | $F(9, 161) = 1.03$; $p = 0.417$ (0.609) |
| Mean speed in weaving sections (m/s) | $F(3, 59) = 0.23$; $p = 0.875$ (0.875) | $F(3, 167) = 1.33$; $p = .267$ (.535) | $F(9, 167) = 0.81$; $p = .609$ (.609) |
| SSQ total severity (log scale) | $F(3, 54) = 2.14$; $p = 0.106$ (0.106) | $F(3, 152) = 1.75$; $p = 0.160$ (0.160) | $F(9, 152) = 0.93$; $p = 0.504$ (0.504) |

Each cell gives $F(df_1, df_2)$ from a linear mixed model (Type III, Kenward–Roger degrees of freedom), the unadjusted p-value and, in parentheses, the multiplicity-adjusted p-value (Holm across the three primary outcomes; Benjamini–Hochberg within each secondary family). For the conflict rate, cells give Wald $\chi^2$ tests (layout, penetration) and a likelihood-ratio test (interaction) from a negative-binomial mixed model.

The penetration effect was explained by leader composition (Table 4, Figure 3B–C). The share of following time spent behind a truck rose from 7.8% at 20% penetration to 47.6% at 80% (exploratory test: $p < 0.001$). In the 179 drives with both car-following and truck-following samples, drivers spent 28.3% of car-following time but only 13.2% of truck-following time at gaps below 1.5 s. Within each leader type, short-gap exposure did not change with penetration (car-following: $p = 0.36$; truck-following: $p = 0.49$). When the drive-level truck-following share was added as a covariate, the penetration effect disappeared ($p = 0.61$), while each additional 10 percentage points of truck-following time was associated with a 1.6 percentage-point reduction in short-gap exposure ($b = -0.158$, $SE = 0.051$, $p = 0.002$).

In this experiment, higher automated-truck penetration lengthened drivers' safety margins because drivers more often followed a truck, behind which they kept longer gaps, and not because they changed how closely they followed a given type of vehicle.

**Table 4. Safety margins and leader composition by automated-truck penetration. Model means are estimated marginal means from Eq. (4); other rows are drive-level means with 95% confidence intervals.**

| Measure | 20% | 40% | 60% | 80% |
|---|---|---|---|---|
| Short time-gap exposure, gap < 1.5 s (model mean [95% CI]) | 0.285 [0.225, 0.345] | 0.268 [0.209, 0.328] | 0.247 [0.187, 0.307] | 0.181 [0.121, 0.241] |
| Gap < 1.0 s (model mean [95% CI]) | 0.139 [0.101, 0.177] | 0.152 [0.114, 0.190] | 0.117 [0.079, 0.155] | 0.077 [0.038, 0.115] |
| Mean time gap when following, s (model mean [95% CI]) | 3.25 [2.91, 3.58] | 3.40 [3.06, 3.73] | 3.39 [3.05, 3.72] | 3.59 [3.25, 3.92] |
| Share of following time behind a truck (mean [95% CI]) | 0.078 [0.039, 0.117] | 0.159 [0.118, 0.201] | 0.255 [0.191, 0.319] | 0.476 [0.395, 0.557] |
| Gap < 1.5 s when following a car (mean [95% CI]) | 0.294 [0.235, 0.353] | 0.310 [0.238, 0.381] | 0.299 [0.234, 0.365] | 0.246 [0.179, 0.314] |
| Gap < 1.5 s when following a truck (mean [95% CI]) | 0.160 [0.055, 0.264] | 0.156 [0.089, 0.223] | 0.107 [0.061, 0.153] | 0.117 [0.067, 0.166] |

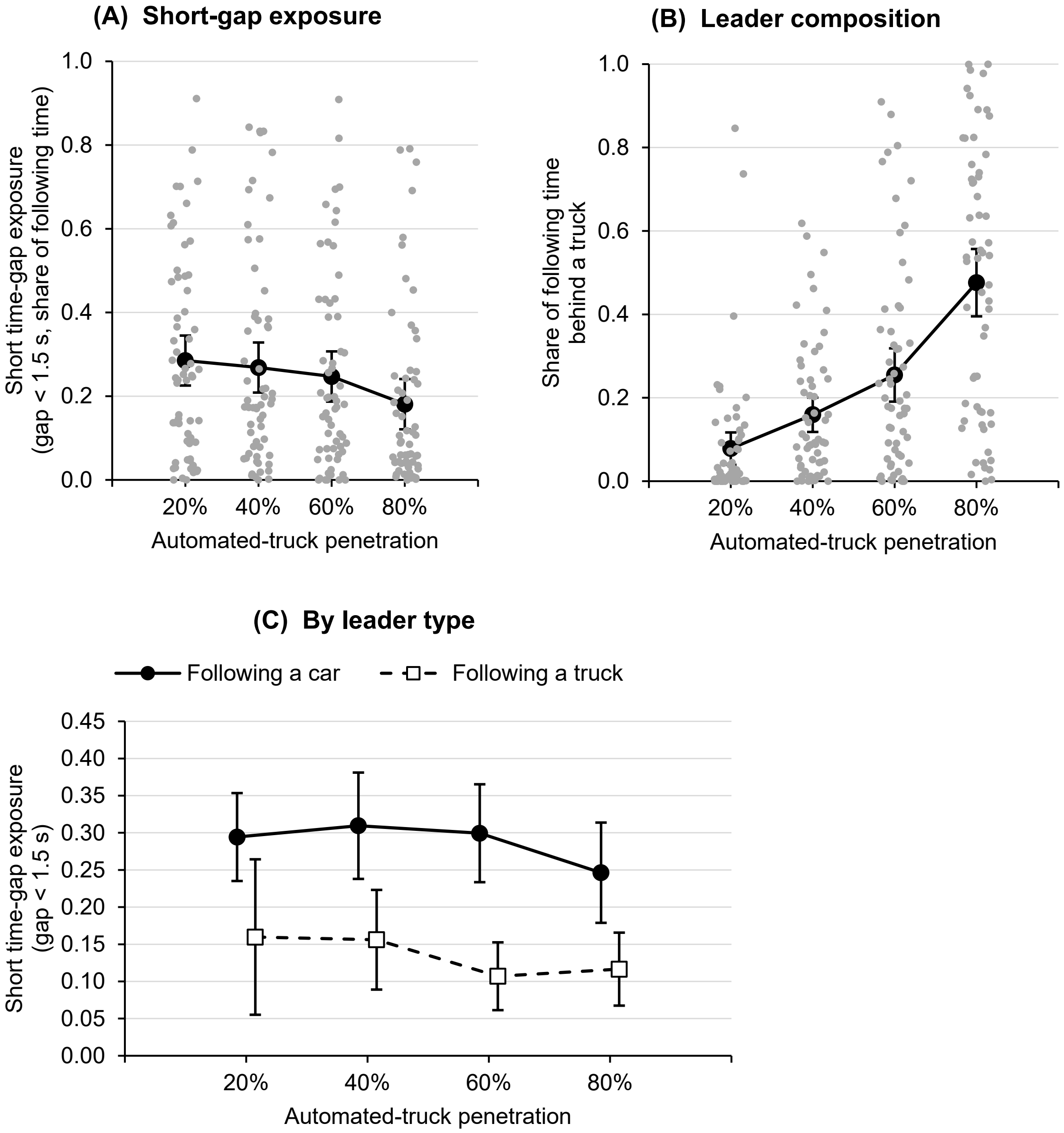

**Figure 3. Car-following safety margins by automated-truck penetration: (A) Short time-gap exposure (estimated marginal means with 95% CI; grey points are individual drives). (B) Share of following time spent behind a truck (means with 95% CI). (C) Short time-gap exposure when following a car and when following a truck (means with 95% CI).**

After the 300-meter pre-merge zones were excluded, 113 rear-end conflicts remained. They occurred in 63 of the 241 drives and involved 39 of the 63 drivers (Table 5). Of these, 100 occurred within the weaving sections, at a median of 218 m downstream of the merge. The retained events were close interactions. The median minimum TTC was 0.61 s (interquartile range 0.08–1.01 s), and the median maximum closing speed was 12.4 m/s. In 70% of events, the participant reduced speed by more than 2 m/s, and 44% coincided with a logged contact. The leader was a truck in 18% of events. The conflict rate averaged 4.0 per 100 km. It did not depend on penetration ($\chi^2(3) = 0.21$, $p = 0.98$; model means 2.9–3.4 per 100 km) or on the layout × penetration interaction (LR $\chi^2(9) = 6.74$, $p = 0.66$). The layout effect was not significant ($\chi^2(3) = 6.98$, $p = 0.073$; BH-adjusted $p = 0.29$). Descriptively, the rate was lowest with the longest AVDL (layout a, 1.5 per 100 km) and highest where the AVDL coincided with the weaving section (layout b, 4.9 per 100 km). The rate ratio of layout (a) to layout (b) was 0.31 for all conflicts (Holm-adjusted $p = 0.062$) and 0.27 for conflicts within the weaving sections (Holm-adjusted $p = 0.047$). The overall layout tests were also not significant for weaving-section conflicts ($\chi^2(3) = 7.21$, $p = 0.065$) or for severe conflicts with minimum TTC ≤ 1.0 s ($\chi^2(3) = 7.67$, $p = 0.053$). With 113 events among 63 drivers, this layout pattern is a hypothesis to be tested rather than a finding.

**Table 5. Rear-end conflicts between the participant and the vehicle ahead by AVDL layout.**

| AVDL layout | Drivers with conflicts / all drivers | Conflicts (in weaving sections) | Rate per 100 km [95% CI] | Median min. TTC (s) | Median max. closing speed (m/s) | Speed reduction > 2 m/s (%) | Logged contact (%) |
|---|---|---|---|---|---|---|---|
| (a) 1800 m | 7 / 16 | 13 (11) | 1.5 [0.7, 3.3] | 0.84 | 11.2 | 69 | 38 |
| (b) 1500 m | 11 / 17 | 43 (41) | 4.9 [2.7, 9.0] | 0.67 | 12.4 | 70 | 44 |
| (c) 1200 m | 11 / 15 | 31 (26) | 4.0 [2.1, 7.6] | 0.47 | 15.0 | 77 | 48 |
| (d) No AVDL | 10 / 15 | 26 (22) | 3.2 [1.7, 6.1] | 0.35 | 10.7 | 62 | 42 |
| All layouts | 39 / 63 | 113 (100) | – | 0.61 | 12.4 | 70 | 44 |

Note: Rates are estimated marginal means from the negative binomial mixed model with 95% confidence intervals. The other columns describe the retained events.

## 4.2 Travel time and speed

Measured over identical road sections, the four layouts produced virtually the same travel times (Figure 4). The mean time to traverse a 1.5-km weaving section was 58.0 s (95% CI 55.5–60.6) in layout (a), 57.8 s (55.3–60.4) in (b), 59.4 s (56.7–62.1) in (c) and 58.6 s (56.0–61.3) without an AVDL, $F(3, 59) = 0.29$, $p = .84$. All pairwise differences between layouts lay within −5.2 s and +4.5 s (95% CIs), so layout effects larger than about 9% of weaving-section travel time can be excluded. Full-route travel time (484–499 s) and weaving-section speed (25.6–26.1 m/s) also did not differ by layout or penetration, and adjusting for gender and age did not change the layout result ($p = 0.81$). Travel time decreased by 0.9 s per weaving section with each successive drive (SE 0.25, $p < 0.001$), consistent with familiarisation with the route.

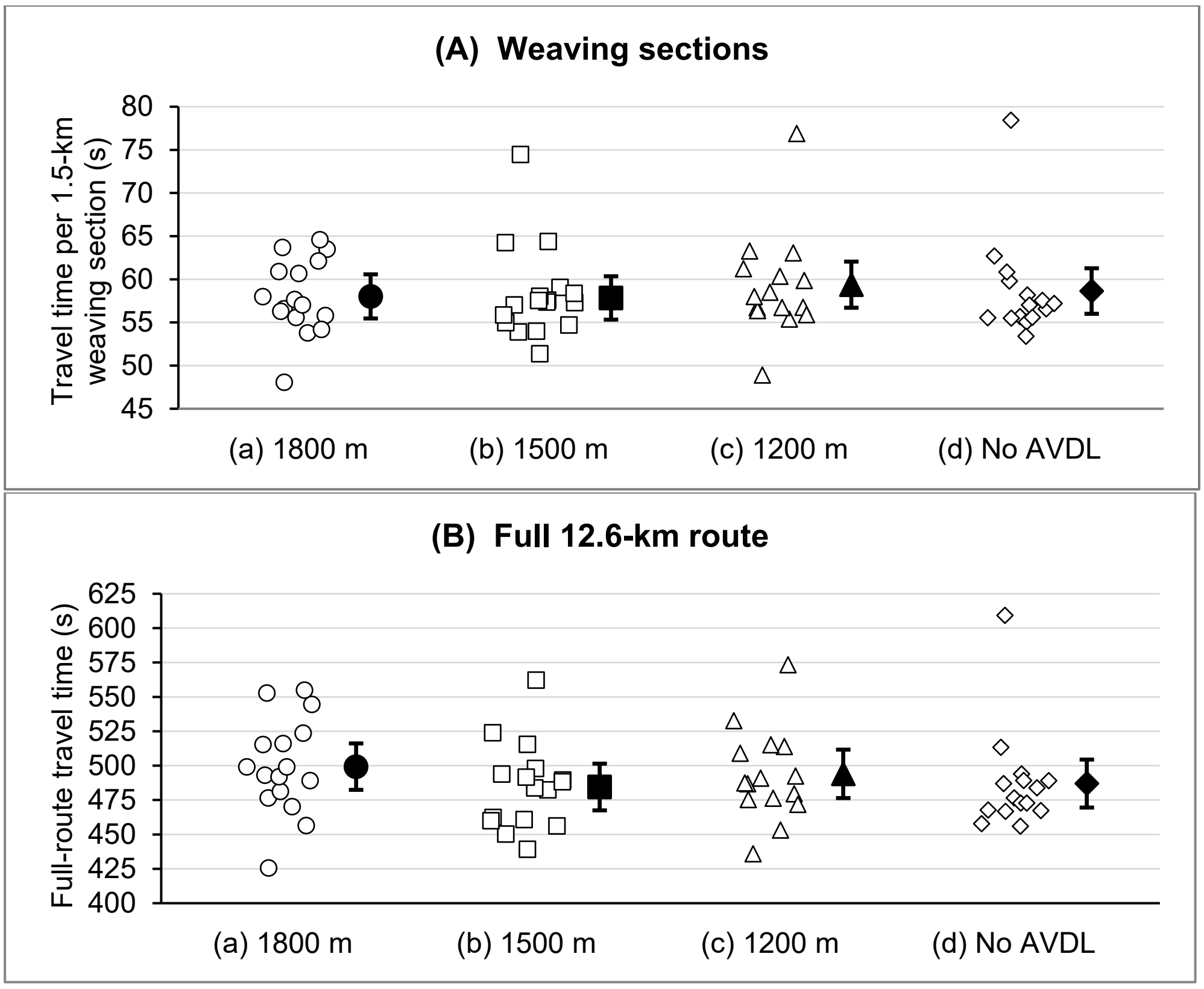

**Figure 4. Travel time by AVDL layout measured over identical road sections: (A) Mean time per 1.5-km weaving section; (B) full-route travel time.** Point symbols are participant means; black markers are estimated marginal means with 95% CI.

## 4.3 Workload, simulator sickness, heart rate and gaze

Raw NASA-TLX scores were lowest in layout (a) (estimated mean 29.4, 95% CI 21.4–37.3) and highest in layout (c) (43.7, 35.7–51.7), with layouts (b) (40.1) and (d) (36.6) in between (Figure 5A). The layout effect was not significant ($p = 0.081$; Holm-adjusted $p = 0.24$). The largest pairwise difference, between (a) and (c), was 14.4 points (Holm-adjusted $p = 0.083$). The confidence intervals for pairwise layout differences (−25.7 to +18.2 points) are wide, so moderate-to-large layout effects on workload can neither be confirmed nor excluded. Adjusting for gender and age left this conclusion unchanged ($p = 0.067$; layout (a) 28.5 and (c) 43.7). Penetration did not affect workload ($p = 0.79$). Among the subscales, physical demand and effort were nominally higher in layouts (b) and (c) (uncorrected $p = 0.028$ and 0.031), but no subscale effect survived false-discovery-rate correction (all adjusted $p \geq 0.09$).

Simulator sickness was low (median total severity 3.7–16.8 across layouts; Figure 5B) and did not differ by layout ($p = 0.11$) or penetration ($p = 0.16$). Mean heart rate (80–84 bpm) showed no layout or penetration effects.

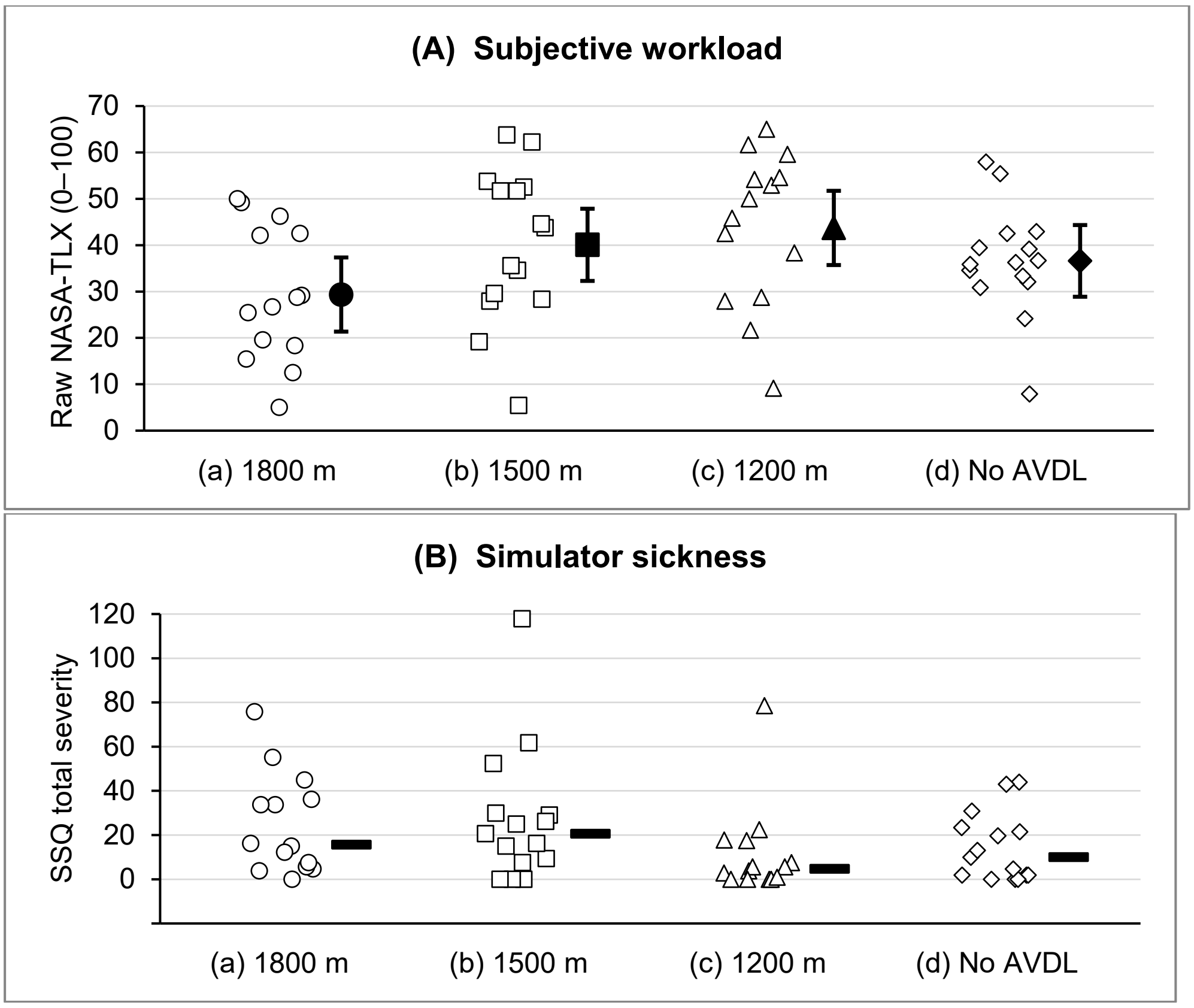

**Figure 5. Subjective workload and simulator sickness by AVDL layout: (A) Raw NASA-TLX (0–100); black markers are estimated marginal means with 95% CI. (B) SSQ total severity (Kennedy et al., 1993)**; black bars are medians. Point symbols are participant means.

The gaze ray hit an object in 72% of samples on average. In 72 of the 241 drives (30%), fewer than half of the samples had a hit, which left 169 drives by 50 drivers for the gaze analysis. In these drives, vehicles received 41% of gaze hits, road and roadside 45% and unknown targets 13% (Figure 6A).

Total gaze on vehicles did not change with penetration (43.0% of hits at 20% and 40.0% at 80%; $p = 0.97$), but its composition did (all three BH-adjusted $p < 0.001$): gaze on cars fell from 35.4% to 11.3% of hits ($p < 0.001$); gaze on trucks rose from 7.6% to 28.6% ($p < 0.001$); trucks' share of vehicle gaze therefore rose from 21% to 74% ($p < 0.001$).

Gaze on road and roadside and on unknown targets did not change ($p = 0.98$ and 0.23). Transitions between AOIs shifted in the same direction. At 20% penetration, 52% of transitions out of road and roadside went to a car. At 80%, 58% went to a truck (Figure 6B–C).

Within drivers, trucks' share of vehicle gaze tracked the share of following time spent behind a truck ($r = 0.76$ across 167 drives, after removing each driver's mean). However, the penetration effect on truck gaze remained when the truck-following share was controlled ($p < 0.001$). Trucks therefore drew gaze even when they were not the vehicle being followed.

Penetration did not affect the other gaze measures. Vehicle visit rate (14–18 per minute), mean vehicle visit duration (2.5–3.4 s) and AOI switching rate (30–35 per minute) did not change (all p

≥ 0.26). Stationary and transition entropy rose slightly (normalised entropy 0.58 to 0.69 and 0.67 to 0.74; p = 0.039 and 0.028), but neither effect survived correction (BH-adjusted p = 0.069 and 0.055). The horizontal gaze-angle measures differed by penetration (entropy p = 0.004, BH-adjusted p = 0.015) without a monotonic trend. No gaze measure differed by layout (all p ≥ 0.32), and using all 241 drives gave the same conclusions (Appendix, Table A2).

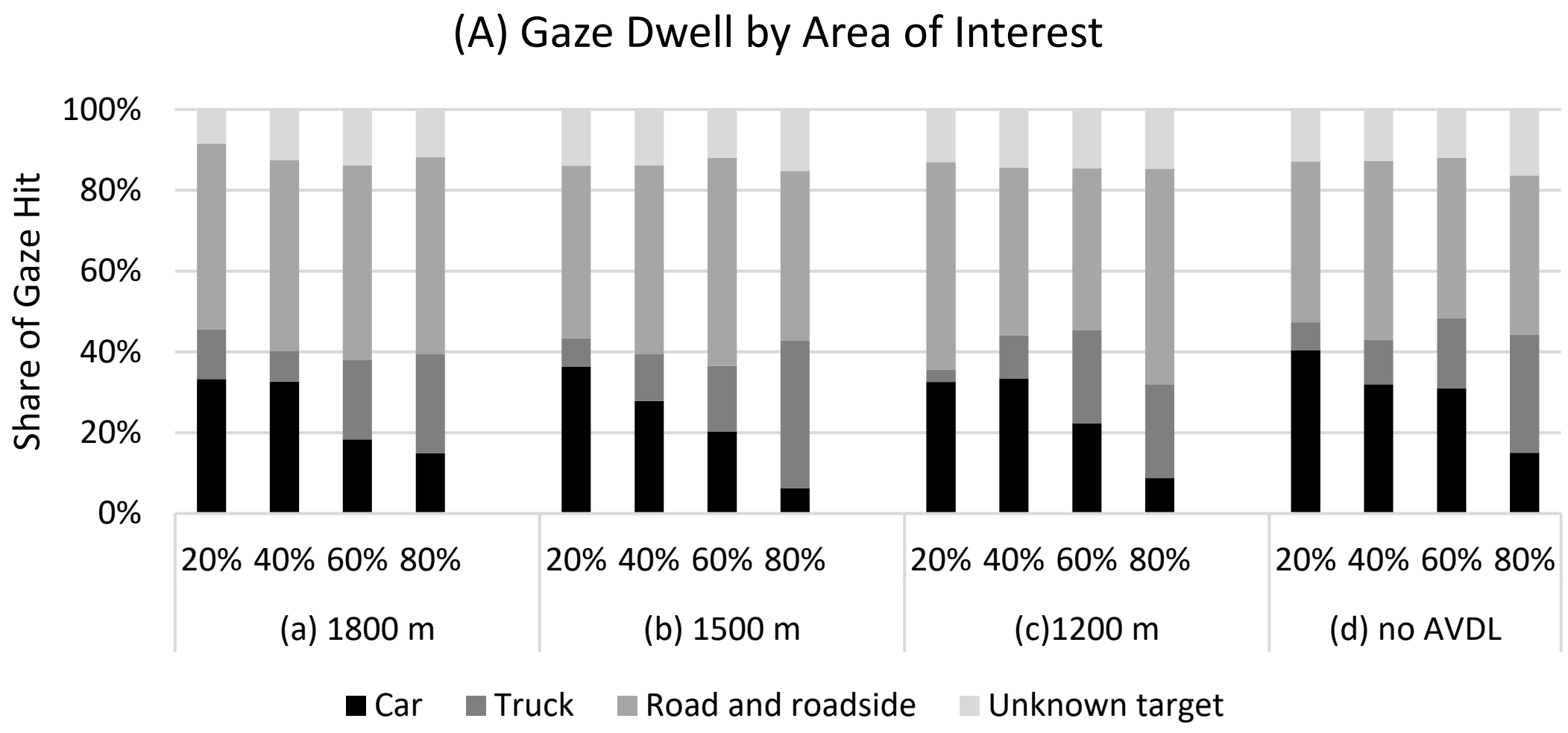

**(B) Transition at 20% of Penetration**

| From AOI \ To AOI | Car | Truck | Road | Unknown |
|---|---|---|---|---|
| Car | – | 0.15 | 0.65 | 0.20 |
| Truck | 0.32 | – | 0.51 | 0.17 |
| Road | 0.52 | 0.17 | – | 0.31 |
| Unknown | 0.35 | 0.12 | 0.53 | – |

**(C) Transition at 80% of Penetration**

| From AOI \ To AOI | Car | Truck | Road | Unknown |
|---|---|---|---|---|
| Car | – | 0.36 | 0.47 | 0.17 |
| Truck | 0.13 | – | 0.63 | 0.24 |
| Road | 0.15 | 0.58 | – | 0.27 |
| Unknown | 0.10 | 0.43 | 0.47 | – |

**Figure 6. Gaze areas of interest (AOIs): (A) Share of gaze hits on cars, trucks, road and roadside, and unknown targets (invisible walls and boxes in the scene) by AVDL layout and automated-truck penetration**; bars are means over drives in which at least half of the samples hit an object. **(B, C) Probability that the next AOI visited is the column AOI, given the current row AOI, pooled over the same drives at 20% and 80% penetration.**

## 4.4 Between-driver heterogeneity

For each of the main outcomes, stable differences between drivers accounted for a much larger share of variance than the experimental conditions (Figure 7). The participant random intercept accounted for 52–61% of the total variance in the main driving measures, 66–67% in workload and simulator sickness, and 77% in mean heart rate, whereas the fixed effects of layout, penetration and serial position together accounted for 2–10%. Equivalently, the intraclass correlations (conditional on fixed effects) were 0.57 for short-gap exposure, 0.62 for mean time gap, 0.56 for weaving-section travel time, 0.75 for NASA-TLX and 0.83 for mean heart rate. Gaze composition was the exception. Penetration accounted for 30% of the variance in gaze on trucks and 43% in trucks' share of vehicle gaze, more than the between-driver differences (19% and 15%), because it changed which vehicles were in view.

Drivers differed in their overall level but not detectably in how they responded to penetration. The variance of driver-specific penetration slopes was not significant for any primary outcome (short-gap exposure $p = 0.36$, weaving-section travel time $p = 0.059$, NASA-TLX $p = 0.22$; Holm-adjusted $p \geq 0.18$) or for the other safety measures (mean time gap $p = 0.083$; BH-adjusted $p \geq 0.17$). It was significant for full-route travel time (BH-adjusted $p = 0.041$) and for some exploratory outcomes, including gaze on cars and on trucks (BH-adjusted $p = 0.007$ and 0.018) and mean heart rate (BH-adjusted $p = 0.026$). Drivers therefore differed in how far their gaze shifted towards trucks as penetration rose, but not in how their car-following margins changed.

Across the four drives, drivers became faster and followed more closely on average, which is consistent with familiarisation. Per drive, weaving-section travel time fell by 0.9 s (95% CI 0.4–1.4), mean speed in the weaving sections rose by 0.32 m/s (0.13–0.52), the mean time gap shortened by 0.13 s (0.03–0.23) and mean heart rate fell by 1.6 bpm (1.1–2.1). Short-gap exposure ($p = 0.20$), NASA-TLX ($p = 0.56$) and the desired time gap in close following (ratio 0.99 per drive, $p = 0.80$) did not change. The serial-position covariate separates these trends from the effects of penetration.

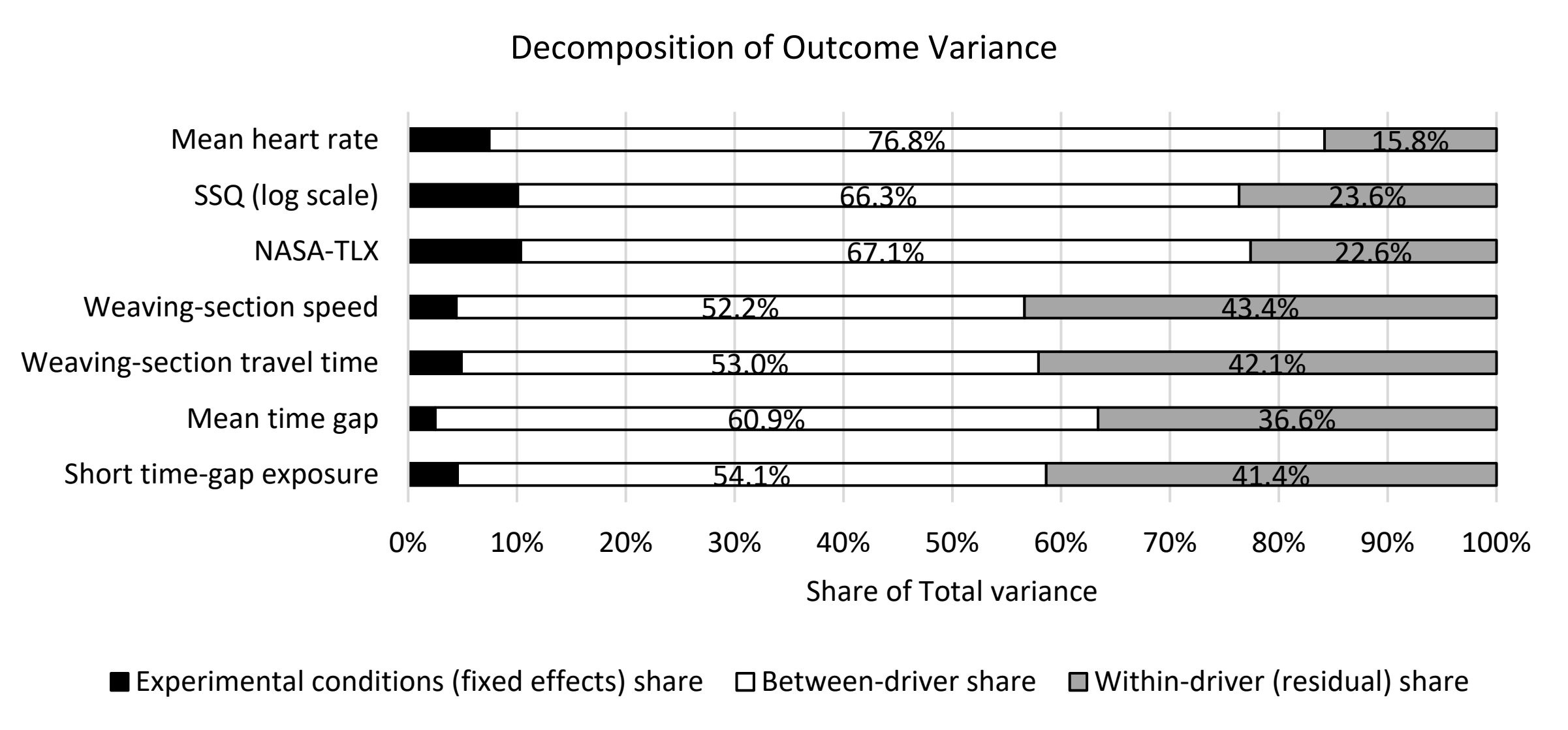

**Figure 7. Decomposition of outcome variance into experimental conditions (fixed effects), stable between-driver differences (participant random intercept) and within-driver residual variation (Nakagawa and Schielzeth, 2013).** Labels give the between-driver share.

## 4.5 Leader-type-dependent car-following model

The calibration used 487 close-following episodes (6.6 h) from 56 drivers: 391 behind cars (5.9 h) and 96 behind trucks (0.6 h, 28 drivers). The episodes covered motorway following at 14–34 m/s (1st–99th percentiles of speed), with no stop-and-go traffic. Within an episode, the leader's speed varied by a median of 2.5 m/s, and decelerations stronger than 3 m/s$^2$ (1-s means) occupied 0.7% of close-following time for leaders and 1.1% for participants. Episodes ended at any jump in forward range, such as when another vehicle cut in or the leader changed lane, so cut-ins were not part of the calibration data. The common parameters were similar across the four specifications (Table 6): $v_0$ = 34.9 m/s, $a$ = 3.6–3.9 m/s$^2$ (close to its upper bound of 4 m/s$^2$), $b$ = 1.5–1.6 m/s$^2$ and $s_0$ = 6.6–6.9 m. Because $s_0$ and $T$ trade off at the speeds observed, they are best read together as an equilibrium time gap at the median operating speed of 26.3 m/s. Under the driver model, this equilibrium time gap ranged from 1.0 s to 2.4 s across drivers (10th–90th percentiles; median 1.6 s). Letting the desired time gap differ between drivers reduced the held-out spacing error from 0.352 to 0.339, and the error was lower for 27 of 49 drivers (Wilcoxon p = 0.057).

Leader type mattered less. The common truck multiplier was θ = 1.11, which raised the median equilibrium time gap from 1.58 s behind cars to 1.73 s behind trucks.

In the episode-level mixed model, the ratio of desired time gaps behind trucks and cars was 1.13 (95% CI 0.93–1.37; p = 0.22). The ratio was 1.02 when episodes near the pre-merge zones were excluded, and 1.20 (0.97–1.48; p = 0.085) among the 27 drivers who followed both leader types. Drivers did not differ in their response to truck leaders, and the variance of driver-specific leader-type effects was estimated at zero (p = 0.50). Neither leader-dependent specification predicted held-out episodes better than the driver model (NRMSE 0.340 with a common multiplier and 0.346 with driver-specific truck gaps, against 0.339; Table 6). Truck-led held-out episodes were predicted no better with driver-specific than with population parameters.

Most of the variation in the desired time gap lay within drivers. Of the episode-level variance not explained by the fixed effects, 16% lay between drivers, 29% between drives of the same driver and 55% between episodes within a drive. Part of the episode-level share is estimation error in episodes as short as 15 s, and part is situational. The desired time gap was 6.6% shorter for each 1 m/s of episode speed (95% CI 3.6–9.5%; p < .001), which may partly reflect the trade-off between $s_0$ and $T$. Adding speed left the leader-type ratio almost unchanged (1.14, 0.94–1.38).

Once drivers were following closely, they followed trucks much as they followed cars: the median time gap in close-following episodes was 1.61 s behind trucks and 1.57 s behind cars. The difference lay in how often drivers closed up. They spent 39% of their truck-following time at a time gap below 3 s, against 56% of their car-following time (difference −16 percentage points, 95% CI −21 to −12; p < 0.001).

Table A3 in Appendix repeats the car-following analysis of Section 4.5 under alternative calibration objectives, trajectory processing, episode selections and model forms. The relative conclusions did not depend on the calibration choices, but the absolute parameter values did (Appendix, Table A3). Across the ten calibrations, letting the desired time gap differ between drivers lowered the pooled held-out error every time, although the advantage for individual drivers was not significant in any ($p \geq 0.057$). Leader-dependent specifications lowered the held-out error by at most 0.001, θ ranged from 0.97 to 1.29 and the episode-level truck/car ratio from 0.98 to 1.23. The ratio was significant only when episodes were restricted to median time gaps below 2 s (1.23, 95% CI 1.02–1.49, $p = 0.033$). The share of the variance of $T$ lying between drivers ranged from 9% to 37%. It was largest, and the share between episodes smallest (38%), when only episodes of at least 30 s were used, which suggests that much of the episode-to-episode variation is estimation error in short episodes. The common parameters, by contrast, changed with the objective, the smoothing, the episode selection and the model form. Across the variants, $b$ ranged from 0.7 to 5.4 m/s$^2$ and $s_0$ from 1.4 to 8.3 m. With a reaction delay, $s_0$ fell to 1.4–2.0 m, but the held-out error did not improve.

**Table 6. Calibrated IDM specifications for close following (487 episodes, 56 drivers).**

| Specification | Desired time gap T | $v_0$ | $a$ | $b$ | $s_0$ | NRMSE, all episodes | Held-out NRMSE: all | Held-out: car-led | Held-out: truck-led |
|---|---|---|---|---|---|---|---|---|---|
| Population | One value for all drivers (0.92 s) | 34.9 | 3.60 | 1.50 | 6.61 | 0.336 | 0.352 | 0.361 | 0.215 |
| Driver | One value per driver (56) | 34.9 | 3.92 | 1.55 | 6.88 | 0.285 | 0.339 | 0.347 | 0.232 |
| Leader-type | One value per driver, × θ = 1.11 behind trucks | 34.9 | 3.89 | 1.63 | 6.56 | 0.285 | 0.340 | 0.348 | 0.244 |
| Driver × leader type | One value per driver and leader type (83) | 34.9 | 3.92 | 1.57 | 6.88 | 0.281 | 0.346 | 0.349 | 0.321 |

Note: Common parameters are from calibration on all episodes ($v_0$ in m/s, a and b in m/s$^2$, $s_0$ in m). NRMSE is the normalised root-mean-square error of the simulated gap. Held-out values are for the episodes not used in calibration (221 episodes from 49 drivers, including 39 truck-led episodes from 19 drivers), with parameters calibrated on the remaining episodes.

# 5 Discussion

## 5.1 Automated trucks change who human drivers follow

The clearest result is that a higher share of automated trucks lengthened human drivers' car-following margins, and that this effect operated entirely through leader composition. As penetration rose from 20% to 80%, the share of following time spent behind a truck increased about sixfold. Drivers kept substantially longer gaps behind trucks than behind cars. Within each leader type, however, gaps did not change with penetration, and the penetration effect vanished once the truck-following share was controlled. This contrasts with reports of headway adaptation after exposure to platoons with short internal gaps (Gouy et al., 2014). One explanation is that the automated trucks in our scenarios kept relatively long internal gaps and travelled in short groups. Another is that our design measured following behaviour over whole drives rather than immediately after close interaction with a platoon. The penetration effect did not extend to rear-end conflicts, whose rate was similar at all penetration levels. Conflicts arose when drivers closed quickly on a slower leader, mostly within the weaving sections, rather than during steady following, so longer steady-state gaps behind trucks need not reduce them.

Drivers' gaze followed the same compositional shift. About 41% of gaze hits fell on vehicles at every penetration level, but trucks' share of that gaze rose from 21% to 74%, in step with the share of time spent behind a truck. Higher penetration therefore redistributed gaze among vehicles rather than drawing more of it to traffic, and the rate and duration of vehicle visits did not change. Truck gaze also rose beyond what truck-following explained, which is consistent with trucks' larger visual size and their presence in adjacent lanes. It has to be noted that the ray-cast records the object in the gaze direction rather than a fixation. Gaze on a vehicle shows that the vehicle lay in the driver's line of sight, not that it was attended to, and fixation-level eye tracking would be needed to separate the two.

The direction of the leader-type difference also needs care. Naturalistic and instrumented-vehicle studies found that drivers followed light trucks and vans more closely than cars (Sayer et al., 2000; Brackstone et al., 2009), the opposite of what we observed for 15-m automated trucks. Several features of the simulator could favour longer gaps behind trucks. These include the underestimation of distances in virtual reality (Renner et al., 2013), the large visual size of a truck rear in a head-mounted display, the trucks' speed relative to the participant (which may have led drivers to overtake rather than follow closely), and drivers' unfamiliarity with automated heavy vehicles. The compositional mechanism itself does not depend on the sign: whenever drivers follow trucks at systematically different gaps from cars, changing the truck share will shift aggregate safety margins without any change in behaviour. The direction and size of the effect in real traffic, however, must be established with field data.

The finding has two implications for traffic modelling. First, mixed-traffic simulations that represent human-driven vehicles with a single car-following parameter set, or with parameters that vary with AV penetration, will misattribute this effect. The calibrated car-following model (Section 4.5) shows where the leader-type dependence lies. Once drivers followed closely, their dynamics behind trucks and cars were similar. A common truck multiplier on the desired time gap was not reliably different from one and did not improve prediction. This differs from Reddy et al. (2026), who found shorter safe time headways and jam spacings behind AVs than behind human-driven cars in a simulator. In their experiment, the leader's type and driving style were manipulated in a dedicated following task. Here, leader type varied naturally within drives, and trucks differed from cars in size as well as in automation.

What differed was how often drivers closed up behind a truck. In simulation, the effect is therefore better represented in the decisions that precede close following than in the close-following parameters themselves. These decisions include the gap at which drivers settle behind a large or slower vehicle and their propensity to change lanes or overtake. Differences between drivers in the desired time gap were much larger than the leader-type effect, so heterogeneous parameters matter more than leader-specific ones (Section 5.3). Second, the safety gain may have a capacity counterpart. If the longer gaps behind trucks persisted in the dense traffic near capacity, which this experiment did not include, they would reduce the flow that human-driven occupied lanes can carry. The net effect of automated truck penetration on human-driven vehicle lane capacity would then depend on how often human-driven vehicles end up behind trucks. That in turn depends on the lane management policy, including whether trucks are confined to an AVDL. Calibrating these decisions will require field trajectories in which the type of the vehicle ahead is identified.

## 5.2 AVDL layout near ramps had little effect on human drivers

Across safety margins, travel time and speed, the three AVDL layouts and the no-AVDL baseline were statistically indistinguishable, and the confidence intervals exclude layout effects on weaving-section travel time larger than about 5 s per section. In the range tested here, moving the AVDL entry and exit by ±150 m relative to the ramp gores therefore did not measurably change how human drivers moved through the weaving section. Two cautions apply. First, the layout comparison was between participants and powered only for large effects ($f \geq 0.35$), so moderate effects on safety margins or workload cannot be excluded. Second, the human driver was one vehicle among simulated traffic, and system-level outcomes such as capacity and conflict frequency among simulated vehicles were not the focus of this study. The participants' own rear-end conflict rate also did not differ significantly by layout. Descriptively, it was lowest with the longest AVDL (layout a) and highest with the AVDL that coincided with the weaving section (layout b). One possible explanation is that in layout (a) automated vehicles enter and leave the dedicated lane outside the ramp influence area, so fewer of them change lanes in front of human drivers within it.

Two features of the data argue for caution. The pattern rests on 113 events, and the layout (a) group included proportionally more women and more participants aged 30 or over (Table 1). Testing the explanation would require more conflicts and, ideally, a within-participant comparison of layouts. The results are consistent with the view that AVDL layout should be chosen mainly on the basis of traffic-operational criteria. Within the range tested, the behaviour of individual drivers did not discriminate between layouts.

Workload scores were descriptively higher with the shorter AVDLs (layouts b and c) than with the longest AVDL (layout a), but this pattern was not significant after accounting for between-participant variation and multiple testing. A plausible hypothesis is that an AVDL whose access points fall inside the weaving section requires drivers to anticipate more lane changes by automated vehicles within the section. Because the layout effect was estimated between groups of 14–15 participants, and because individual differences dominated the workload scores, this hypothesis requires a within-participant test.

## 5.3 Heterogeneity among drivers

Stable differences between drivers explained more than half of the variance in each of the main outcomes, roughly six to twenty-four times as much as the experimental conditions. Two consequences follow. For experimental design, between-participant comparisons of infrastructure layouts require much larger samples than within-participant manipulations, or designs in which each participant experiences several layouts. For simulation-based evaluation of AVDLs, a single calibrated driver is a poor representation of the population. To focus on the safety margins, heterogeneous parameter distributions will be needed (Ossen and Hoogendoorn, 2011; Saifuzzaman and Zheng, 2014). The car-following model points the same way: equilibrium time gaps ranged from 1.0 s to 2.4 s between the 10th and 90th percentile drivers, and letting them differ between drivers improved held-out predictions, whereas letting them differ between leader types did not. Within drivers, however, the desired time gap varied more from one following episode to the next than it differed between drivers, partly with speed. This is consistent with trajectory studies showing that a driver's car-following parameters change within a trip (Taylor et al., 2015), but also this could be due to the familiarisation with the driving simulator. Averaged over a drive, this variation gives way to the stable differences between drivers seen in the drive-level outcomes. Drivers also did not differ detectably in how their following responded to penetration or to truck leaders. Within the conditions observed in this study, simulation models therefore need distributions of car-following parameters across drivers and variation within drivers over time, but perhaps not driver-specific responses to automated trucks.

### 5.4 Limitations

The study has several limitations. The layout comparison was between participants with 15–17 participants per group. Random assignment produced groups that did not differ significantly in gender or age, but chance imbalances of this size are expected with small groups. Participants were mostly young university students, and driving experience was not recorded. Penetration order was varied but only approximately counterbalanced, although serial position was controlled statistically. A single demand level, road geometry and set of automated-truck behaviours were tested, and the automated trucks' control logic and realised shares were not logged at the vehicle level. Leader type was inferred from a simulator variable rather than recorded directly; the inference was unambiguous for almost all samples, but it applies only to the vehicle detected straight ahead. Every truck was automated, so the leader-type effect cannot be separated from the trucks' size, automated control and platooning. The car-following model rests on 96 truck-led close-following episodes from 28 drivers, about half of them at 80% penetration. Longer truck-following episodes were fragmented, because at long distances, the forward range jumped between parts of a truck. The IDM has no reaction delay, and the narrow speed range prevented separate estimation of the standstill gap and the desired time gap. The maximum acceleration was also close to its bound. The parameters are therefore effective values for close following at motorway speed in this simulator rather than transferable estimates.

Car-following models estimated in simulators generally reproduce field behaviour in structure but not in parameter values (Papadimitriou and Choudhury, 2017; Paschalidis et al., 2021). The parameters in Table 6 are therefore best used for relative comparisons, between leader types and between drivers, and would need to be combined with field trajectories before being used in field simulation. Only the participant's vehicle was logged. Conflicts among simulated vehicles, lane-change and crossing conflicts, and post-encroachment time therefore could not be computed, and a full SSAM analysis was not possible. The rear-end conflict measure covers only the participant and the vehicle directly ahead. Future simulator studies should log the trajectories of all vehicles near the participant in a format that SSAM can read.

Heart-rate data were 1-Hz averages and could not support heart-rate-variability analysis. Gaze was logged at 10 Hz without fixation detection. In 30% of drives, fewer than half of the samples hit an object, and 14% of hits fell on invisible walls or boxes whose target is unknown. The road surface had no collider, so gaze on the road could not be separated from gaze on the roadside. The physiological and visual-attention results are therefore exploratory. Finally, the study was conducted in a fixed-base VR simulator with relatively short drives, and its findings require confirmation with naturalistic data from corridors where dedicated lanes and automated trucks are deployed.

The car-following model has further limits of scope. It describes behaviour in this simulator experiment, and its parameters should not be read as representative of on-road drivers. The episodes covered motorway following at 14–34 m/s, without stop-and-go traffic and with few hard decelerations. The model has therefore not been validated for congested, stop-and-go or safety-critical conditions, in which simulator-calibrated models can underestimate drivers' braking (Berghaus et al., 2021). The leaders were simulated vehicles whose speed varied little within an episode, and episodes ended when another vehicle cut in, and only the vehicle directly ahead was logged. The unpredictable braking, cut-ins and reactions to several vehicles ahead that occur in everyday traffic are therefore not represented. Drivers also became faster and followed more closely across the session, and their awareness of taking part in an experiment may have affected their behaviour. The serial-position covariate controls for the first of these but not the second. The held-out comparison tests prediction within the experiment, not validity in real traffic, which requires naturalistic driving records or road experiments with instrumented vehicles (Sangster et al., 2013; Huang et al., 2018). Finally, the conclusions rest on one model family. The relative conclusions were largely unchanged under alternative objectives, smoothing, episode selections and reaction delays (Appendix, Table A3), but this cannot show that the IDM's structure is uniquely correct, and the absolute parameter values changed with these choices. A population parameter set built from these driver-level estimates would also need to account for correlations between parameters (Treiber and Kesting, 2013; James et al., 2019).

## 6 Conclusions

This driving-simulator study examined how AVDL layout near motorway ramps and the share of automated trucks affect the human drivers who share the road. Four conclusions follow.

First, a higher share of automated trucks reduced human drivers' exposure to short car-following gaps, from 28.5% to 18.1% of following time. This reduction arose because drivers more often followed a truck, behind which they kept longer gaps in this simulator, not because they changed their following behaviour. Drivers' gaze on vehicles shifted from cars to trucks in the same way, while total gaze on vehicles stayed constant. Because naturalistic studies report the opposite gap difference for light trucks, the sign of this effect in real traffic needs field confirmation.

Second, shifting the AVDL entry and exit by 150 m relative to the ramp gores, or omitting the AVDL, did not measurably change human drivers' safety margins, rear-end conflict rate, travel time or speed, and its effect on workload was not significant.

Third, a calibrated car-following model located the leader-type effect in how often drivers closed up behind trucks rather than in how they followed once they had. In close following, the truck multiplier on the desired time gap was small (1.11–1.13), uncertain and did not improve prediction. These parameters describe close following at motorway speeds in this simulator experiment; they have not been validated against field data or for stop-and-go or emergency situations.

Fourth, stable differences between drivers outweighed all experimental effects, and driver-specific time gaps improved the model where leader-specific ones did not. Drivers did not differ detectably in how their following responded to penetration or to truck leaders, and within a drive their desired time gap varied more from episode to episode than it differed between drivers. Models used to evaluate AVDLs and automated-truck deployment should therefore represent human drivers as heterogeneous. They should represent the effect of automated trucks through drivers' choices of whether and how closely to follow them, rather than through a single leader-type adjustment to car-following parameters.

## Acknowledgements

The authors acknowledge TRACSLab, UNSW's driving-simulator laboratory. This work was supported in part by the 2024 University of New South Wales and Huazhong University of Science and Technology Strategic Partnership Research Seed Fund: Global Research and Impact Program (GRIP).

## Disclosure statement

The authors report that there are no competing interests to declare.

## Author contributions (CRediT)

Minhao Chen: Software, Investigation, Data curation, Formal analysis, Visualisation, Writing – original draft. Zehao Jiang: Conceptualisation, Methodology, Writing – review and editing. Julius Secadiningrat: Software, Data curation. Elnaz Irannezhad: Conceptualisation, Methodology, Software, Investigation, Formal analysis, Visualisation, Supervision, Funding acquisition, Writing – original draft, Writing – review and editing.

## Use of generative AI

The authors used Claude (Anthropic) to assist with language editing and proofreading. The authors reviewed and verified all text and take full responsibility for the content.

## Data availability statement

The de-identified simulator logs, questionnaire responses and heart-rate exports are openly available at https://github.com/E-Nezhad/AVDL.

## Ethics statement

The study was approved by the University of New South Wales Human Research Ethics Advisory Panel (HREAP) Executive (approval no. iRECS8743). All participants provided written informed consent.

## Appendix: Exploratory outcomes

**Table A1. Linear mixed-model tests for exploratory outcomes, formatted as in Table 3.**

| Outcome | N (drivers/ drives) | Layout | Penetration | Layout × penetration |
|---|---|---|---|---|
| Simulator throughput (log) | 62/239 | F(3, 58) = 0.54; p = .658 (.974) | F(3, 164) = 17.22; p < .001 (< .001) | F(9, 164) = 1.03; p = .419 (.419) |
| Mental demand | 58/225 | F(3, 54) = 1.70; p = .177 (.266) | F(3, 155) = 0.49; p = .692 (.915) | F(9, 154) = 0.74; p = .674 (.944) |
| Physical demand | 58/225 | F(3, 54) = 3.27; p = .028 (.092) | F(3, 155) = 0.39; p = .762 (.915) | F(9, 155) = 0.38; p = .944 (.944) |
| Temporal demand | 58/225 | F(3, 54) = 0.89; p = .451 (.451) | F(3, 155) = 0.10; p = .962 (.962) | F(9, 155) = 1.61; p = .116 (.427) |
| Performance (reversed) | 58/225 | F(3, 54) = 1.70; p = .177 (.266) | F(3, 155) = 0.75; p = .523 (.915) | F(9, 155) = 1.10; p = .364 (.727) |
| Effort | 58/225 | F(3, 54) = 3.19; p = .031 (.092) | F(3, 155) = 1.43; p = .235 (.705) | F(9, 155) = 0.41; p = .929 (.944) |
| Frustration | 58/225 | F(3, 54) = 1.05; p = .380 (.451) | F(3, 155) = 2.57; p = .057 (.340) | F(9, 155) = 1.53; p = .142 (.427) |
| Mean HR (bpm) | 56/213 | F(3, 52) = 0.75; p = .529 (.595) | F(3, 145) = 0.54; p = .658 (.658) | F(9, 145) = 1.28; p = .252 (.452) |
| Within-drive SD of HR | 56/213 | F(3, 52) = 0.64; p = .595 (.595) | F(3, 146) = 0.82; p = .483 (.658) | F(9, 146) = 0.99; p = .452 (.452) |
| Gaze dwell on cars (share of hits) | 50/169 | F(3, 45) = 0.54; p = .657 (.892) | F(3, 110) = 18.01; p < .001 (< .001) | F(9, 110) = 0.37; p = .948 (.961) |
| Gaze dwell on trucks (share of hits) | 50/169 | F(3, 43) = 0.21; p = .890 (.892) | F(3, 113) = 27.73; p < .001 (< .001) | F(9, 113) = 1.15; p = .337 (.961) |
| Gaze dwell on all vehicles (share of hits) | 50/169 | F(3, 45) = 0.26; p = .857 (.892) | F(3, 108) = 0.07; p = .974 (.983) | F(9, 108) = 0.85; p = .573 (.961) |
| Truck share of vehicle gaze | 50/167 | F(3, 43) = 1.20; p = .322 (.892) | F(3, 112) = 48.56; p < .001 (< .001) | F(9, 112) = 1.10; p = .370 (.961) |
| Gaze dwell on road and roadside (share of hits) | 50/169 | F(3, 46) = 0.22; p = .885 (.892) | F(3, 108) = 0.06; p = .983 (.983) | F(9, 108) = 1.17; p = .320 (.961) |
| Gaze on unknown targets (share of hits) | 50/169 | F(3, 45) = 0.22; p = .883 (.892) | F(3, 110) = 1.47; p = .226 (.352) | F(9, 110) = 0.74; p = .668 (.961) |
| Vehicle gaze visits per minute | 50/169 | F(3, 44) = 0.76; p = .520 (.892) | F(3, 111) = 1.36; p = .258 (.361) | F(9, 111) = 0.67; p = .731 (.961) |
| Mean vehicle gaze visit duration (s, log) | 50/167 | F(3, 46) = 0.21; p = .892 (.892) | F(3, 106) = 0.73; p = .538 (.628) | F(9, 106) = 0.64; p = .759 (.961) |
| AOI switches per minute | 50/169 | F(3, 45) = 0.60; p = .621 (.892) | F(3, 110) = 1.05; p = .372 (.474) | F(9, 110) = 0.59; p = .803 (.961) |
| Stationary gaze entropy (normalised) | 50/169 | F(3, 45) = 0.61; p = .612 (.892) | F(3, 111) = 2.88; p = .039 (.069) | F(9, 111) = 0.34; p = .961 (.961) |

| Outcome | N (drivers/ drives) | Layout | Penetration | Layout × penetration |
|---|---|---|---|---|
| Gaze transition entropy (normalised) | 50/169 | F(3, 45) = 0.74; p = .533 (.892) | F(3, 111) = 3.16; p = .028 (.055) | F(9, 111) = 0.51; p = .862 (.961) |
| Horizontal gaze-angle entropy (normalised) | 50/169 | F(3, 46) = 0.28; p = .842 (.892) | F(3, 107) = 4.67; p = .004 (.015) | F(9, 107) = 0.84; p = .582 (.961) |
| Share of gaze in densest 5% of angle bins | 50/169 | F(3, 46) = 0.36; p = .779 (.892) | F(3, 107) = 3.65; p = .015 (.041) | F(9, 107) = 0.73; p = .677 (.961) |
| Horizontal gaze-angle SD (degrees, log) | 50/169 | F(3, 45) = 0.26; p = .855 (.892) | F(3, 111) = 3.52; p = .018 (.041) | F(9, 111) = 0.65; p = .753 (.961) |
| Share of following time behind a truck | 63/241 | F(3, 58) = 1.43; p = .243 (.728) | F(3, 168) = 38.56; p < .001 (< .001) | F(9, 168) = 1.02; p = .424 (.637) |
| Gap <1.5 s share when following a car | 63/237 | F(3, 59) = 0.17; p = .916 (.916) | F(3, 163) = 1.08; p = .360 (.493) | F(9, 163) = 1.29; p = .245 (.637) |
| Gap <1.5 s share when following a truck | 62/183 | F(3, 59) = 0.48; p = .699 (.916) | F(3, 124) = 0.81; p = .493 (.493) | F(9, 123) = 0.72; p = .693 (.693) |
| Logged contact events (any collider) | 63/241 | F(3, 59) = 0.07; p = .974 (.974) | F(3, 167) = 1.22; p = .306 (.306) | F(9, 167) = 2.62; p = .007 (.015) |

Note: Adjusted p-values (in parentheses) use the Benjamini–Hochberg procedure within each family (NASA-TLX subscales; physiology; gaze; mechanism; system outputs).

**Table A2. Logged gaze objects, their identity in the simulator build and the assigned area of interest.**

| Logged object name | What it is in the simulator build | Share of gaze hits (%) | Area of interest |
|---|---|---|---|
| Terrain | Ground, including the road surface (the road mesh has no collider) | 32.9 | Road and roadside |
| NPC_LaneChange_Target | Invisible 2 × 2 × 5 m trigger box enclosing each simulated car | 20.4 | Car |
| HAV_LaneChange_Target | Invisible 3 × 3 × 13 m trigger box enclosing each truck | 12.8 | Truck |
| treemain | Trees | 6.6 | Road and roadside |
| Lane ExitTrigger | Invisible 30 m wide, 15 m high wall across the carriageway, 240 m before the end of each weaving section | 4.9 | Unknown target |
| AV_MergeTrigger | Invisible 30 m wide, 15 m high wall across the carriageway, 220 m before each merge | 4.8 | Unknown target |
| AVDL | Invisible 30 × 15 × 1500 m box at each weaving section (position varies with layout) | 3.9 | Unknown target |
| SM_concrete_block | Concrete barriers | 3.6 | Road and roadside |
| LaneChangeCollider | Invisible 2 × 2 × 5 m trigger box enclosing each merging automated car | 3.4 | Car |
| Cube* (fixed location) | Parts of overhead sign gantries | 2.7 | Road and roadside |
| Cube* (moving) | Hidden collider on semi-trailers | 0.5 | Truck |
| trailer, Cab, Chassis, SemiTruck, AttachmentPoint | Visible truck parts | 2.3 | Truck |
| NPC_HUMAN-DRIVEN VEHICLE, NPC_AV, NPC_Merge_AV | Visible car bodies | 0.6 | Car |
| Collider | Moving collider of unidentified type | 0.5 | Unknown target |
| Pole, Rectangular Vertical | Poles and road signs | 0.1 | Road and roadside |

* Objects named “Cube” were either parts of overhead sign gantries, which do not move, or hidden colliders on semi-trailers. A Cube hit was assigned to road and roadside if its location, on a 2 m × 5 m grid, was hit in at least five drives, and to trucks otherwise; 84% of Cube hits were at such fixed locations. No object was hit in 27.9% of all samples.

**Table A3. Robustness of the car-following conclusions.**

| Variant | Episodes (drivers) | θ | Held-out error: P / A / C / B | Driver model better than population | Truck/car ratio of T (95% CI) | Variance of T: drivers / drives / episodes (%) |
|---|---|---|---|---|---|---|
| Baseline | 487 (56) | 1.11 | 0.352 / 0.339 / 0.340 / 0.346 | 27 of 49 | 1.13 (0.93–1.37) | 16 / 29 / 55 |
| Objective: speed NRMSE | 487 (56) | 1.07 | 0.157 / 0.157 / 0.157 / 0.157 | 27 of 49 | 1.14 (0.97–1.34) | 17 / 7 / 76 |
| Objective: spacing and speed combined | 487 (56) | 1.12 | 0.272 / 0.265 / 0.265 / 0.269 | 27 of 49 | 1.14 (0.94–1.38) | 16 / 29 / 54 |
| Smoothing over 0.5 s | 487 (56) | 1.13 | 0.354 / 0.342 / 0.342 / 0.351 | 27 of 49 | 1.11 (0.94–1.31) | 15 / 35 / 49 |
| Smoothing over 2.1 s | 487 (56) | 1.11 | 0.348 / 0.337 / 0.338 / 0.344 | 29 of 49 | 1.10 (0.91–1.34) | 16 / 29 / 55 |
| Median time gap < 2 s | 344 (48) | 1.29 | 0.349 / 0.331 / 0.330 / 0.334 | 19 of 40 | 1.23 (1.02–1.49) | 14 / 27 / 58 |
| Episodes ≥ 30 s | 237 (47) | 1.06 | 0.344 / 0.333 / 0.337 / 0.335 | 23 of 39 | 0.98 (0.67–1.42) | 37 / 25 / 38 |
| Away from pre-merge zones | 299 (54) | 0.97 | 0.334 / 0.329 / 0.329 / 0.355 | 25 of 43 | 1.00 (0.80–1.24) | 9 / 30 / 61 |
| Reaction delay 0.5 s | 487 (56) | 1.12 | 0.356 / 0.347 / 0.348 / 0.354 | 27 of 49 | 1.09 (0.92–1.28) | 13 / 35 / 52 |
| Reaction delay 1.0 s | 487 (56) | 1.13 | 0.361 / 0.352 / 0.353 / 0.360 | 27 of 49 | 1.18 (0.99–1.40) | 16 / 26 / 59 |

Note: P, A, C and B denote the population, driver, leader-type and driver × leader-type specifications. Held-out errors are in each variant's own objective, so they can be compared between specifications within a row but not between rows; the speed error hardly distinguished between specifications, which is consistent with speed-based measures being less sensitive than gap-based ones (Treiber and Kesting, 2013). The truck/car ratio and the variance shares are from the episode-level mixed model for $\log T$ with random intercepts for driver and drive. Across the variants, the common parameters of the driver model ranged from 3.0 to 4.0 m/s$^2$ for $a$, 0.7 to 5.4 m/s$^2$ for $b$ and 1.4 to 8.3 m for $s_0$. Episodes (drivers): episodes calibrated, and drivers contributing. θ: truck multiplier of the leader-type specification. Driver model better than population: drivers whose held-out error was lower with driver-specific than with population

parameters. Variance of T: shares of the variance of the episode-level desired time gap between drivers, between drives of the same driver and between episodes within a drive.